\documentclass[twocolumn, twocolappendix]{aastex631}

\usepackage{color}
\usepackage{epsfig}
\usepackage{soul}
\usepackage{float}
\usepackage{enumitem}
\usepackage{appendix}
\usepackage{wrapfig}
\usepackage{graphicx}
\usepackage{hyperref}
\usepackage{chemfig}
\usepackage{longtable}
\usepackage{tabularx}
\usepackage{graphicx}
\usepackage{supertabular,booktabs}
\usepackage[version=3]{mhchem}
\hypersetup{
    colorlinks=true,
    linkcolor=blue,
    filecolor=magenta,
    citecolor=blue,
    urlcolor=blue,
    pdftitle={Overleaf Example},
    pdfpagemode=FullScreen,
    }
\usepackage{natbib}

\def\aap{A\&A}
\def\apjs{ApJS}

\usepackage{blindtext}
\usepackage{lipsum}
\usepackage{afterpage}

\newsavebox{\splitcaptionbox}
\newlength{\splitcaptionheight}
\makeatletter
\long\def\@makesplitcaption#1#2{
  \if@stage@final
    \vskip0.7\abovecaptionskip
    \addtolength{\splitcaptionheight}{-0.7\abovecaptionskip}%
  \else
    \vskip\abovecaptionskip\goodbreak
    \addtolength{\splitcaptionheight}{-\abovecaptionskip}%
  \fi
  \setbox\splitcaptionbox=\vbox{\interlinepenalty 0
    \reset@font\small{\bfseries#1.} #2\par}%
  \vsplit\splitcaptionbox to \splitcaptionheight
  \global\setbox\splitcaptionbox=\box\splitcaptionbox
  \if@cop@home\ifonline\ifnum\csname c@\@captype\endcsname=1 
    \immediate\write\@auxout{\string\gdef\string\@num\@captype{}}%
    \hypertarget{\@captype}{}%
  \fi\fi\fi}

\newcommand{\splitcaption}[3][\empty]
 {\setlength{\splitcaptionheight}{#3}%
  \let\cop@makecaption=\@makecaption
  \let\@makecaption=\@makesplitcaption
  \ifx\empty#1\relax
    \caption{#2}%
  \else
    \caption[#1]{#2}%
  \fi
  \let\@makecaption=\cop@makecaption}
\makeatother

\newcommand{\mergecaption}{\ifdim\ht\splitcaptionbox>0pt
  \begin{figure}[tp]
  \box\splitcaptionbox
  \end{figure}
\fi}

\begin{document}

\title{High sensitivity molecular line observations towards the L1544 pre-stellar core challenge current models \footnote{Based on observations carried out with the IRAM 30m Telescope. IRAM is supported by INSU/CNRS (France), MPG (Germany) and IGN (Spain)}}

\author{J.~Ferrer Asensio}
\affiliation{Centre for Astrochemical Studies, Max-Planck-Institut f\"ur extraterrestrische Physik, Giessenbachstr. 1, 85748 Garching, Germany}
\affiliation{RIKEN Pioneering Research Institute, Wako-shi, Saitama, 351-0106, Japan}

\author{S. S.~Jensen}
\affiliation{Centre for Astrochemical Studies, Max-Planck-Institut f\"ur extraterrestrische Physik, Giessenbachstr. 1, 85748 Garching, Germany}

\author{S.~Spezzano}
\affiliation{Centre for Astrochemical Studies, Max-Planck-Institut f\"ur extraterrestrische Physik, Giessenbachstr. 1, 85748 Garching, Germany}

\author{P.~Caselli}
\affiliation{Centre for Astrochemical Studies, Max-Planck-Institut f\"ur extraterrestrische Physik, Giessenbachstr. 1, 85748 Garching, Germany}

\author{F. O.~Alves}
\affiliation{Centre for Astrochemical Studies, Max-Planck-Institut f\"ur extraterrestrische Physik, Giessenbachstr. 1, 85748 Garching, Germany}

\author{O.~Sipil{\"a}}
\affiliation{Centre for Astrochemical Studies, Max-Planck-Institut f\"ur extraterrestrische Physik, Giessenbachstr. 1, 85748 Garching, Germany}

\author{E.~Redaelli}
\affiliation{Centre for Astrochemical Studies, Max-Planck-Institut f\"ur extraterrestrische Physik, Giessenbachstr. 1, 85748 Garching, Germany} \affiliation{European Southern Observatory, Karl-Schwarzschild-Stra{\ss}e 2, 85748 Garching, Germany}



\begin{abstract}

The increased sensitivity and spectral resolution of observed spectra towards the pre-stellar core L1544 are challenging the current physical and chemical models. With the aim of further constraining the structure of L1544 as well as assessing the completeness of chemical networks, we turn to radiative transfer modelling of observed molecular lines towards this source. We obtained high-sensitivity and high-spectral resolution observations of \ce{HCO+} (J = 1 - 0), CS (J = 2 - 1), C$^{34}$S (J = 2 - 1), \ce{H2CO} (J$_{K_{a},K_{c}}$ = 2$_{1,2}$ - 1$_{1,1}$), c-C$_{3}$H$_{2}$ (J$_{K_{a},K_{c}}$ = 2$_{1,2}$ - 1$_{0,1}$), SO (N,J = 2,3 - 1,2) and $^{34}$SO (N,J = 2,3 - 1,2) with the IRAM 30m telescope towards the dust peak of L1544. A non-Local Thermodynamic Equilibrium radiative transfer code is coupled to the Markov Chain Monte Carlo method to model the observations. We find that with just one transition for each isotope, the modelling cannot find a global minimum that fits the observations. The derived fractional abundance profiles are compared to those computed with chemical models. The observed transitions trace gas components with distinct dynamics at different distances along the radius of the core. Moreover, the results evidence a poor reproduction of sulphur chemistry by chemical modelling and stresses the need to include a more consistent S-depletion process to accurately reproduce the S-chemistry towards dense cores.

\end{abstract}



\section{Introduction} \label{sec:intro}

Pre-stellar cores are high-density regions formed from the fragmentation of filamentary structures that constitute molecular clouds \citep{palmeirim:13, hacar:22}. While pre-stellar cores are in the early stages of gravitational contraction, they have not yet formed a protostar. These objects are characterised by high densities ($n_{H_{2}} > 10^{5}$ cm$^{-3}$) and low temperatures (T $<10$ K) at their centre \citep{keto:08}. The study of the physical, chemical and dynamical properties of these cores and their surroundings helps us gain a comprehensive understanding of the initial conditions that lead to core collapse and, therefore, to understand the earliest stages of the star formation process \citep{andre:14}.\ 

Molecular spectra have proven to be valuable tools for characterizing dense cloud cores. Molecular line intensities, widths, and profiles are influenced by the physical , chemical and dynamical conditions along the path of the emitted radiation. The information embedded within observed spectral lines is extracted through a comprehensive radiative transfer modelling of the emission accounting for the physical structure of the observed source, as well as the molecular abundance distribution computed with an astrochemical code (e.g., \citealt{caselli:02, sohn:07, keto:15, redaelli:19a,ferrerasensio:22}). Upon the fitting of the observed molecular transitions with a specific chemical model we can additionally gain information on the source's evolutionary status.\

L1544 is a well-known pre-stellar core in the Taurus Molecular Cloud at a distance of 170 pc \citep{galli:19}. Extensive research has focused on the study of its physical and chemical properties, revealing it is centrally concentrated, with low central temperatures and signs of contraction motions \citep{ward:99, crapsi:05, crapsi:07}.\ 

With access to increasingly high-sensitivity and high-spectral resolution spectra of molecules, we are challenging the current models and knowledge on L1544, learning about its sub-structure and the influence of its surroundings. For example, the analysis of high-spectral resolution observed spectra of CO, H$_{2}$O, N$_{2}$H$^{+}$, HCO$^{+}$ and isotopologues towards this source revealed important information on the structure and kinematics of L1544 including the CO depletion towards the centre of the core and the the ionization fraction across the core \citep{caselli:99, caselli:02b}. Moreover, past studies showed the need for adjusting the velocity profile of the physical model \citep{keto:15} to fit the profile of specific lines \citep{bizzocchi:13, ferrerasensio:22}. In addition, spatially extended abundant molecules such as \ce{HCO+} have shown the need to take into account a diffuse envelope around L1544 to be able to reproduce their molecular profiles \citep{redaelli:19a}. Moreover, \cite{lin:22} found a local gas density enhancement towards L1544 deviating from the commonly assumed spherically symmetric Bonnor Ebert (BE) sphere structure, which could be the result of an accretion flow impact.\ 

In this work, we deepen our comprehension of the physical, chemical and kinematic structure of L1544 complementing past works \citep{tafalla:98, caselli:99, caselli:02, bizzocchi:13, redaelli:19a, ferrerasensio:22, lin:22}, through the radiative transfer modelling of new high-sensitivity and high-spectral resolution observations of molecular species towards this pre-stellar core. To probe the entirety of the core, we observe rare isotopologue transitions, C$^{34}$S (J = 2 - 1) and $^{34}$SO (N,J = 2,3 - 1,2), expected to trace the whole core, and main isotopologue transitions, \ce{HCO+} (J = 1 - 0), CS (J = 2 - 1), \ce{H2CO} (J$_{K_{a},K_{c}}$ = 2$_{1,2}$ - 1$_{1,1}$), c-C$_{3}$H$_{2}$ (J$_{K_{a},K_{c}}$ = 2$_{1,2}$ - 1$_{0,1}$) and SO (N,J = 2,3 - 1,2), which are expected to trace the outer parts of the core as they are optically thick. It is important to note that this is the first time that such high-sensitivity and high-spectral resolution data of these molecular transitions are available for this source. To adjust the pre-stellar core model as a consequence of the new information provided by recent observations \citep{ferrerasensio:22}, we adopt a non-Local Thermodynamic Equilibrium (non-LTE) approach coupled with a Markov Chain Monte Carlo (MCMC) method. This allows us to explore the parameter space of selected physical and chemical model variables to provide the highest-probability model solutions to fit the observations. Here, we assess the molecular fractional abundance profiles computed with the pseudo-time dependent chemical model (pyRate) described in \cite{sipila:15} by comparing them to simple step-abundance profiles found to give the best fits of observed transitions towards L1544.\ 
 
This manuscript is structured as follows: the observations are presented in Section \ref{observations3}, the radiative transfer methodology used to reproduce the observed molecular transitions is described in Section \ref{radiative3}, the modelling results are introduced in Section \ref{results3}, these results are interpreted and placed in the context of past works in Section \ref{discussion3} and we summarize our conclusions in Section \ref{conclusions3}. Additionally, we include further tests in the Appendix.\

\section{Observations}\label{observations3}

\begin{table*}[ht]
\centering
\caption{Observed Molecular Transitions}

\resizebox{\textwidth}{!}{
\begin{tabular}{ lccccccccc } 
\hline\hline
Molecule & Transition & Frequency &  $E_\mathrm{up}$ & $A$ & n$_{crit}^{c}$ @ 10 K & rms & HPBW & Main Beam efficiency & Velocity resolution \\
 & & (MHz) & (K) & (x 10$^{-5}$ s$^{-1}$) & (cm$^{-3}$) & (mK) & (") & (\%) & (km s$^{-1}$) \\
\hline
\ce{HCO+} & $J$ = 1 - 0 & 89188.525 & 4.28 & 4.19 & 1.7e5& 64 & 27.6 & 80 & 0.033 \\
HC$^{17}$O$^{+}$ & $J$ = 1 - 0 & 87057.53 & 4.18 & 3.9 & 1.5e5 & 2.8 & 28.3 & 81 & 0.034 \\
\ce{CS} & $J$ = 2 - 1 & 97980.95 & 7.05 & 1.68 & 3.6e5 & 44 & 25.1 & 79 & 0.029 \\
C$^{34}$S & $J$ = 2 - 1 & 96412.94 & 6.94 & 1.60 & 3.6e5 & 16 & 25.5 & 79 & 0.030 \\
\ce{H2CO} & $J _{K_{a},K_{c}}$ = 2$_{1,2}$ - 1$_{1,1}$ & 140839.50 & 21.92 & 5.30 & 6.8e5 & 61 & 17.5 & 74 & 0.020 \\
c-C$_{3}$H$_{2}$ & $J _{K_{a},K_{c}}$ = 2$_{1,2}$ - 1$_{0,1}$ & 85338.89 & 6.44 & 2.32 & 1.7e5 & 52 & 28.8 & 82 & 0.034 \\
SO & $N,J$ = 2,3 - 1,2 & 99299.87 & 9.22 & 1.12 & 1.1e5 & 16 & 24.7 & 79 & 0.029 \\
$^{34}$SO & $N,J$ = 2,3 - 1,2 & 97715.39$^{b}$ & 9.09 & 1.07 & 1.1e5 & 19 & 25.1 & 79 & 0.030 \\
 \hline
\label{obsdat}
\end{tabular}
}
\tablecomments{ The spectral information for the molecular transitions including the frequency, the upper energy, E$_{up}$, and the Einstein coefficient, $A$, was taken from The Cologne Database for Molecular Spectroscopy (CDMS)\footnote{\url{https://cdms.astro.uni-koeln.de}} (HCO$^{+}$, \citealt{tinti:07}, HC$^{17}$O$^{+}$, \citealt{dore:01b}, CS and C$^{34}$S, \citealt{bogey:81}, \ce{H2CO}, \citealt{brunken:03}, c-C$_{3}$H$_{2}$, \citealt{bogey:86}, SO, \citealt{clark:76} and $^{34}$SO, \citealt{tiemann:74}). $^{b}$ The $^{34}$SO frequency was directly measured in the Center for Astrochemical Studies (CAS) laboratory (Valerio Lattanzi, private communication). $^{c}$ The critical densities of the C$^{34}$S and $^{34}$SO isotopologues have been calculated using the collisional rate coefficients of their main isotopologues because the collisional rate coefficients for these rarer isotopologues are not available.}

\end{table*}

Single-pointing observations of the \ce{HCO+} (J = 1 - 0), CS (J = 2 - 1), C$^{34}$S (J = 2 - 1), \ce{H2CO} (J$_{K_{a},K_{c}}$ = 2$_{1,2}$ - 1$_{1,1}$), c-C$_{3}$H$_{2}$ (J$_{K_{a},K_{c}}$ = 2$_{1,2}$ - 1$_{0,1}$), SO (N,J = 2,3 - 1,2) and $^{34}$SO (N,J = 2,3 - 1,2) rotational transitions are obtained towards the dust peak of L1544 ($\alpha _{2000}$ = 05$^h$04$^m$17$^s$.21,  $\delta _{2000}$ = +25$^\circ$10$'$42$''$.8). These observations were carried out in August and October of 2021, using the IRAM 30m telescope at Pico Veleta, and took a total of 5.7 hours of integration time. The telescope pointing was frequently checked against the nearby bright quasars B0316+413, B0851+202, and B0439+360, depending on the elevation of the source. We used the E090 and E150 bands of the EMIR receiver. The observations were performed in frequency-switching mode. We used the VESPA backend to achieve a spectral resolution of 10~kHz. The spectral velocity resolution for each transition is reported in Table \ref{obsdat}. Both horizontal and vertical polarisations were observed simultaneously. The $T_{sys}$ ranged between 65 and 173 K. The summary of the observed transitions in this project alongside HC$^{17}$O$^{+}$ presented in \cite{ferrerasensio:22} can be found in Table \ref{obsdat}.\

\begin{figure*}[t]
\centering
\includegraphics[width=\textwidth]{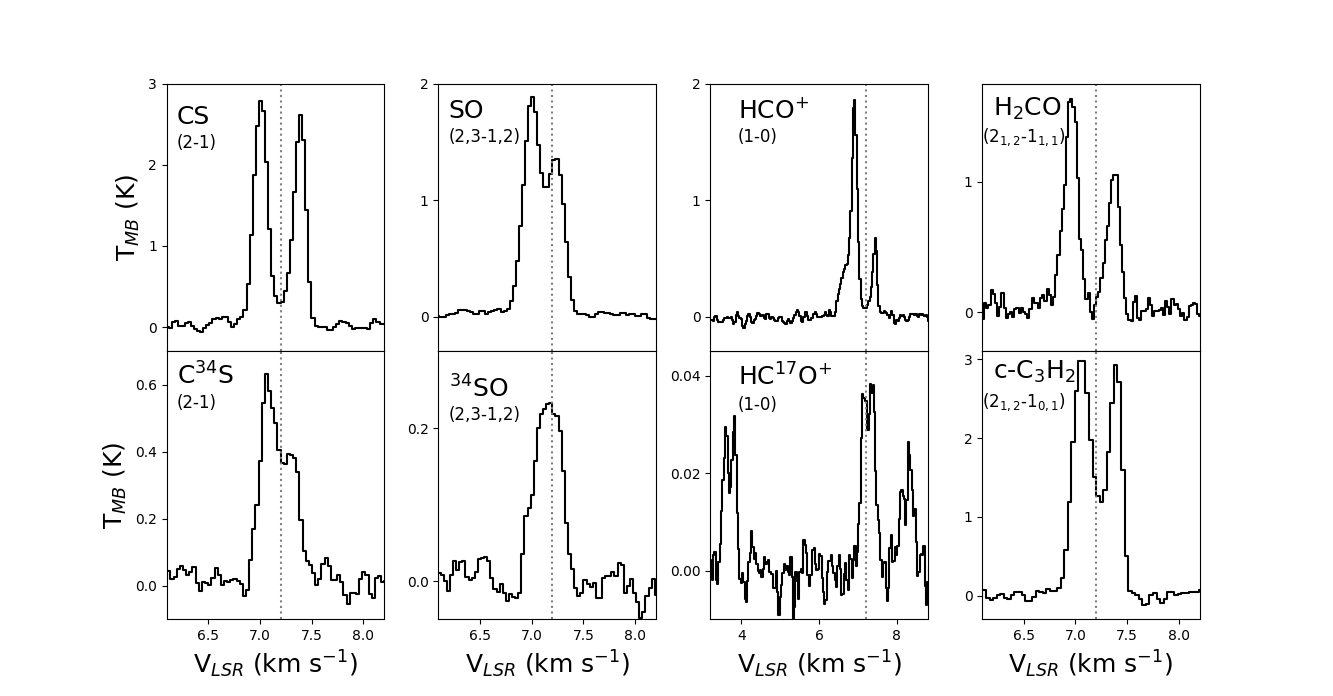}
\caption{Spectra of the molecular lines observed towards the dust peak of L1544. HC$^{17}$O$^{+}$ is included from previous work \citep{ferrerasensio:22}. The vertical dotted grey line represents the LSR velocity of L1544 (7.2 km s$^{-1}$). The misalignment between the SO (2,3 - 1,2) and  c-C$_{3}$H$_{2}$ (2$_{1,2}$ - 1$_{0,1}$) line centres and the LSR of L1544 lays within the laboratory measured transition frequency uncertainties.}
\label{obspap3}
\end{figure*}

The observational data was processed and then averaged with the Continuum and Line Analysis Single-dish Software (\textsc{class}), an application from the \textsc{gildas}\footnote{\url{https://www.iram.fr/IRAMFR/GILDAS/}} software \citep{pety:05}. The resulting spectra, whose intensities have been converted to main beam temperature (see Table \ref{obsdat}), are presented in Figure \ref{obspap3}. 

Signal-to-noise ratio (S/N) peak values are 11 for $^{34}$SO, 13 for the least bright component of \ce{H2CO}, 26 for \ce{HCO+}, 28 for \ce{C$^{34}$S}, 48 for the least bright component of \ce{CS}, 49 for the least bright component of c-C$_{3}$H$_{2}$, and 69 for the least bright component of \ce{SO}.\ 

\newpage

\section{Radiative transfer} \label{radiative3}

In pre-stellar cores, where the molecular hydrogen volume density ($n_{H_{2}}$) and gas temperature ($T_{gas}$) can span between $\gtrsim$10$^{6}$ cm$^{-3}$ and $\sim$ 6 K at the centre and $\lesssim$10$^{2}$ cm$^{-3}$ and $\sim$ 20 K at the edge, we cannot assume LTE conditions throughout the core. The LTE regime is a reasonable assumption when the critical density of the transition ($n_{crit}$= $A_{ul}$/$C_{ul}$ for optically thin lines, where $A_{ul}$ is the Einstein A coefficient and $C_{ul}$ is the collisional coefficient at a specific temperature, and where `$u$' and `$l$' are the upper and lower levels, respectively), is substantially lower than the volume density of the emitting region. Thus, for high $n_{crit}$ transitions ($\geq$ 10$^5$ cm$^{-3}$), LTE applies only in a reduced area of the pre-stellar core centre where the $n_{H_{2}}$ exceeds $n_{crit}$. In non-LTE the energy level populations deviate from the Boltzmann distribution and need to be calculated solving the statistical equilibrium equations, as we describe below.\

\subsection{The LOC radiative transfer code + MCMC} \label{loc3}

\begin{figure*}[t]
\centering
\includegraphics[width=\textwidth]{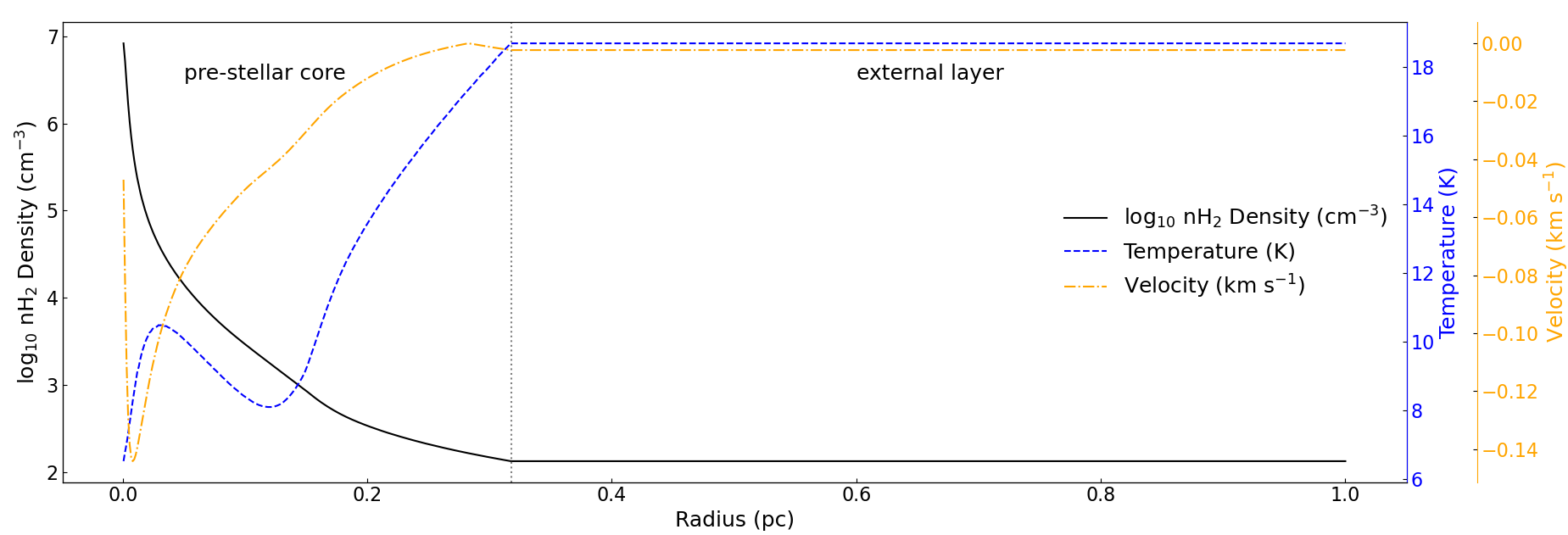}
\caption{1D pre-stellar core physical model profiles used for the radiative transfer modelling. The model described in \cite{keto:15} is plotted between 0 and 0.32 pc. The vertical dotted black line marks the radius at 0.32 pc. The physical profiles between 0.32 and 1 pc correspond to an "external layer" used for some of the transition modelling. The physical parameters in this external layer are constant profiles with values equal to the value at the edge (0.32 pc) of the pre-stellar core model in \cite{keto:15}. The molecular hydrogen column density, $n_{H_{2}}$, is plotted logarithmically with a black solid line. The gas temperature, $T$, is plotted with a blue dashed line. Finally, the velocity profile is plotted with an orange dash-dotted line. }
\label{keto}
\end{figure*}

Radiative transfer modelling of molecular transitions towards L1544 is typically aided by a detailed 1D physical pre-stellar core model presented in \cite{keto:15}. This model has been successful at reproducing many molecular transitions observations towards the dust peak of L1544 \citep{keto:10, keto:101, caselli:12,caselli:17,caselli:22, bizzocchi:13, redaelli:18, redaelli:19a, redaelli:21, redaelli:22b, ferrerasensio:22}. The radial profiles of the pre-stellar core physical model are presented in Figure \ref{keto}. The molecular hydrogen volume density (solid black line) ranges from 8$\times10^{6}$ cm$^{-3}$ at the centre to 1$\times10^{2}$ cm$^{-3}$ at the edge of the core (0.32 pc). The gas temperature (blue dashed line) ranges from 6 K at the centre to 18 K at the edge. Finally, the gas velocity (orange dash and dotted line) ranges from -0.14 km s$^{-1}$ at the velocity peak ($\sim$ 0.01 pc) to -0.01 km s$^{-1}$ at the edge of the core. The negative sign of the velocity indicates that the pre-stellar core model is contracting. \ 

In previous works, the adjustment of physical and chemical model parameters, such as the fractional abundance or the velocity profile scaling, aided the fit and understanding of the nature of some molecular emission towards L1544 \citep{bizzocchi:13, ferrerasensio:22, redaelli:22b}. In this work, we explore the parameter space of specific physical and chemical parameters when performing the radiative transfer modelling of the observations to ensure the best possible fit. With this aim in mind, we combine non-LTE radiative transfer modelling with the Markov Chain Monte Carlo method (MCMC) to calculate probability distributions of the physical and chemical parameters, which produce the closest modelled spectra with respect to the observed spectra.\ 

The radiative transfer modelling of the observed molecular lines is carried out with the Line transfer with OpenCL (LOC) code \citep{juvela:20}. As a first step, LOC solves the statistical equilibrium equations with the radiative and collisional coefficients of the molecule with an accelerated lambda iteration (ALI) \citep{rybicki:91}. Then, the radiative transfer is computed using a 1D model of the source, taking into account the molecular hydrogen volume density, the gas temperature, the velocity of the gas, $v$, the turbulent velocity dispersion,  $\sigma_{turb}$, and the fractional abundance profiles of the desired molecule. The 1D radial profiles of the physical properties are taken from the quasi-statically contracting Bonnor Ebert sphere described in \cite{keto:15}.\  

Additionally to the pre-stellar core physical model, LOC requires information on the fractional abundance profile of the molecules. The fractional abundance profiles for the molecules can be simulated using chemical models. In the past, fractional abundance profiles computed with pyRate were used when performing radiative transfer modelling of L1544 observations. These abundance profiles are simulated using the pre-stellar core physical model described in \cite{keto:15}, which is separated into concentric shells where chemical simulations, based on a chemical network, are conducted. The physical structure of the core is fixed, and the chemistry evolves with time. As one of the objectives is to assess the accurateness of fractional abundance profiles computed with pyRate, we use simple step abundance profiles described in Section \ref{models} for the purposes of comparison.\

The Markov Chain Monte Carlo method, as the name suggests, combines the Markov Chain process with Monte Carlo simulations. The MCMC code EMCEE is used \citep{foreman:13}. The MCMC method involves randomly sampling high-dimensional probability distributions while considering the probabilistic dependence between samples using the Markov Chain. The density estimation for probability distributions is done taking into account both the observations and the model. The value of the parameters that have the lowest $\chi^{2}$ minimization with respect to the observations is the value with the highest likelihood. The $\chi^{2}$ is computed channel wise and then summed over all channels. \ 

First of all, we define the parameters to be explored by the MCMC and a sensible range for their values. Then, initial values, from where the random walk starts, are defined. The MCMC, starting from the initial values, generates a sequence of models using random parameter values. Depending on the $\chi^{2}$ minimization of the resulting model with respect to the observations, the parameters used are accepted or rejected. If they are accepted, these parameter values are used for the next step on the modelling chain. If the parameters are rejected, the previous values in the chain are used.  As a product, we obtain the probability distribution of the parameter values as constrained by the comparison of the produced spectral models with the observed spectra in the form of histograms. In the corner plots, the parameter histograms are arranged to give information on the correlation between the parameters in pairs. \

The models presented in this manuscript were run with 40 walkers for total chains of 10000 steps with a burn-in of 500 steps.  The results correspond to the median values. For a well-constrained model, the median values will converge to the model with the highest probability (lowest $\chi^{2}$). In such cases, the histogram in a corner plot will often resemble a Gaussian profile, depending on the underlying posterior distribution. In cases where the parameter cannot be properly constrained, the histograms are often flat profiles, indicating that the model cannot constrain the parameter, either due to the lack of information or poor assumptions in the model. In the cases where the variable is not constrained, as seen as a flat, instead of Gaussian, histogram in the corner plot, the stated values correspond to the median as the highest probability were not able to be determined. The resulting modelled spectra are convolved with the size of the IRAM 30m beam at each frequency to enable the comparison with the observed spectra. The spectrum simulated by LOC corresponds inherently to a pencil beam, and to account for the beam dilution, the spectrum of the 1D spherically symmetric model is first projected onto a 2D map of the sky. A 2D Gaussian profile, with a FWHM corresponding to the HPBW of the IRAM 30m telescope at the observed frequency, typically extending from –3$\times$FWHM to +3$\times$FWHM, is then convolved with the intensity distribution of this 2D model. A 1D spectrum is subsequently extracted toward the centre of the core in the convolved 2D map. \

\subsection{Models and Parameter Space}\label{models}

Past works required small adjustments of the velocity profile scaling to reproduce some high-sensitivity observations of molecular transitions towards L1544 \citep{bizzocchi:13, ferrerasensio:22}. Thus, for the modelling in this manuscript, we set the velocity profile scaling, $f_{v}$, as a parameter to be explored by the MCMC. Moreover, we include the turbulent velocity dispersion,  $\sigma_{turb}$, as another parameter, as we expect the turbulence to vary in different parts of the core and surroundings. Turbulence is known to decrease within dense cores (e.g. \citealt{fuller:92}; \citealt{goodman:98}; \citealt{pineda:10}; \citealt{choudhury:21}), likely due to the dumping of magneto-hydrodynamic waves (e.g. \citealt{pineda:21}; \citealt{caselli:02}).\ 

Some articles in the literature have also shown the need to scale the fractional abundance profiles computed from pseudo-time-dependent chemical models \citep{redaelli:19a, ferrerasensio:22}. In order to gain insight into the molecular radial distributions directly from observations, we use simple step abundance profiles for the modelling. This choice of abundance profile takes into account the almost total molecular depletion observed towards the centre of the core \citep{caselli:22}, which has an impact on the observed molecular emission.\

\begin{figure}[h]
\centering
\includegraphics[width=9cm]{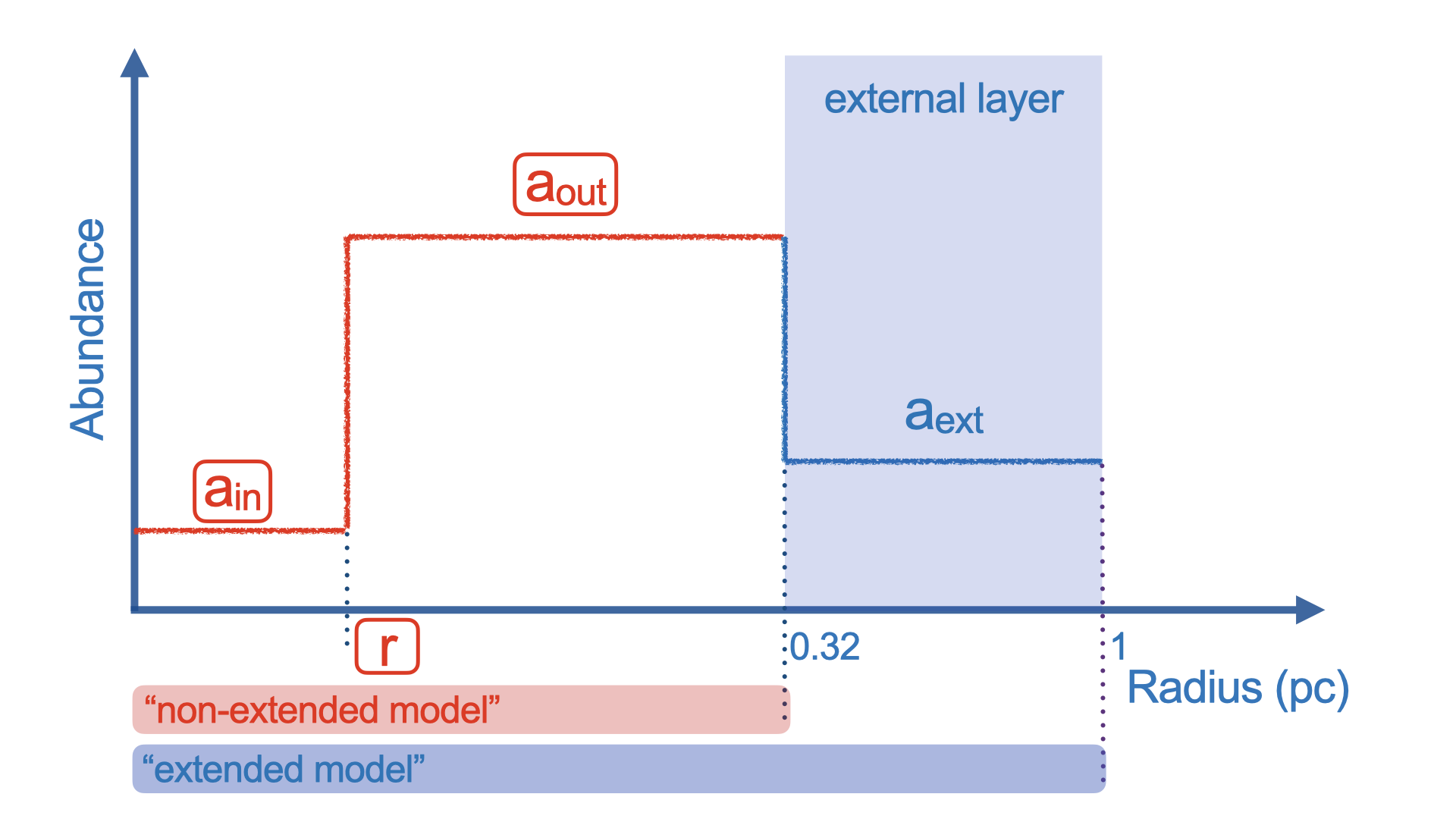}
\caption{Schematics of the abundance profiles used for the radiative transfer modelling. The vertical axis represents the abundance, and the horizontal axis the radius in units of parsec. From the centre of the core (left) to the parameter $r$, we have the inner fractional abundance $a_{in}$ and from $r$ to 0.32 pc, there is the outer fractional abundance $a_{out}$. This constitutes the spatial extent of the "non-extended" model, as indicated by the red stripe at the bottom of the plot. From 0.32 to 1 pc, we have the shaded blue area representing the external layer characterised by an external fractional abundance a$_{ext}$, which we set to the molecular fractional abundance found towards diffuse clouds. Thus, the "extended" model (from the centre to 1 pc) is indicated by the blue stripe at the bottom of the plot. The parameters framed in a box ($a_{in}$, $a_{out}$ and $r$) are those explored by the MCMC. }
\label{abundprof}
\end{figure}

Depending on the nature of the observed molecular transition, we use either a "non-extended" or an "extended" model. The "non-extended" model corresponds to a pre-stellar core physical and chemical model 1D profiles that extend to 0.32 pc as used for the majority of the radiative transfer modelling of molecules towards L1544 \citep{giers:22, ferrerasensio:22}. This model is from \cite{keto:15} but sets the $f_{v}$ and  $\sigma_{turb}$ as parameters to be explored by the MCMC. The molecular abundance profile is constructed with two constant abundance values, one representing the abundance in the inner part of the core, $a_{in}$, and another for the outer part of the core, $a_{out}$. The separation of the inner abundance and outer abundance regions is defined by a radius, $r$. Consequently, the abundance profile has two zones: $a_{in}$ between 0 and $r$ pc and $a_{out}$ between $r$ and 0.32 pc (red stripe in Figure \ref{abundprof}). \ 

The ground state spectra of some abundant molecules, such as HCO$^{+}$,  trace also less dense gas around the pre-stellar core \citep{redaelli:22b}. Thus, for molecules expected to be present in this surrounding extended structure, we adopt the ``extended" model. The extended model consists of the non-extended model structure (0 - 0.32 pc) with an additional layer spanning from 0.32 to 1 pc. The physical parameter profiles in this extended layer are assumed to be constant with the values given at the edge (0.32 pc) of the \cite{keto:15} model. The $f_{v}$ parameter scales the velocity in the entire model from 0 to 1 pc. The abundance is kept constant in this external layer and is set to the observed value towards diffuse clouds. Thus, for the extended model, the abundance profile is defined by $a_{in}$ (between 0 and $r$ pc), $a_{out}$ (between $r$ and 0.32 pc), and diffuse cloud observed abundance $a_{ext}$ (between 0.32 and 1 pc) (blue stripe in Figure \ref{abundprof}). From this point on, the spectra resulting from the non-extended model are plotted with a solid red line, and the spectra resulting from the extended model are plotted with a dash-dotted blue line. \ 

\begin{table*}[ht]
\begin{center}
\caption{LOC + MCMC Parameter Results}

\resizebox{\textwidth}{!}{\begin{tabular}{ lccccc } 
\hline\hline
Transition & a$_{in}$ &  a$_{out}$ & $r$ & f$_{v}$ &  $\sigma_{turb}$ \\
 & & & (au) &  & (km s$^{-1}$) \\
\hline

 C$^{34}$S (2 - 1)  &  1.84$^{+39.46}_{-1.83}$$\times$10$^{-12}$  &  1.82$^{+0.56}_{-1.23}$$\times$10$^{-9}$  &  7697.42$^{+1048.52}_{-1355.66}$  &  0.93$^{+0.41}_{-0.14}$  &  0.10$^{+0.01}_{-0.03}$  \\
 $^{34}$SO (2,3 - 1,2)  &  1.16$^{+56.20}_{-0.03}$$\times$10$^{-13}$  &  3.72$^{+6.79}_{-0.29}$$\times$10$^{-9}$  &  14075.36$^{+2561.94}_{-6273.92}$  &  1.06$^{+0.15}_{-0.25}$  &  0.11$^{+0.02}_{-0.02}$  \\
 HC$^{17}$O$^{+}$ (1 - 0)  &  1.90$^{+6.10}_{-1.87}$$\times$10$^{-13}$  &  4.87$^{+20.66}_{-4.52}$$\times$10$^{-11}$  &  9172.02$^{+7063.61}_{-5658.26}$  &  1.45$^{+0.21}_{-0.13}$  &  0.07$^{+0.01}_{-0.02}$   \\
 CS (2 - 1) &  2.21$^{+2.1}_{-2.11}$$\times$10$^{-10}$  &  1.15$^{+1.21}_{-1.09}$$\times$10$^{-7}$  &  15827.86$^{+305.357}_{-290.997}$  &  1.99$^{+0.00}_{-0.01}$  &  0.01$^{+0.00}_{-0.01}$   \\

 SO (2,3 - 1,2)  &  4.57$^{+3001.51}_{-4.52}$$\times$10$^{-13}$  &  8.93$^{+0.11}_{-7.18}$$\times$10$^{-9}$  &  8013.68$^{+2869.05}_{-145.76}$  &  0.83$^{+0.55}_{-0.02}$  &  0.11$^{+0.00}_{-0.04}$ \\
 
 c-C$_{3}$H$_{2}$ (2$_{1,2}$ - 1$_{0,1}$)  &  1.16$^{+0.05}_{-1.13}$$\times$10$^{-9}$  &  2.33$^{+0.11}_{-0.52}$$\times$10$^{-9}$  &  19408.84$^{+505.24}_{-16032.30}$  &  0.56$^{+0.05}_{-0.06}$  &  0.12$^{+0.00}_{-0.03}$   \\
 c-C$_{3}$H$_{2}$ (2$_{1,2}$ - 1$_{0,1}$) extended  &  3.69$^{+1070.30}_{-3.68}$$\times$10$^{-12}$  &  1.56$^{+0.08}_{-0.14}$$\times$10$^{-9}$  &  1161.61$^{+544.80}_{-493.39}$  &  0.79$^{+0.13}_{-0.08}$  &  0.11$^{+0.00}_{-0.01}$   \\
 HCO$^{+}$ (1 - 0) extended  &  7.85$^{+1439}_{-0.10}$$\times$10$^{-13}$  &  2.00$^{+2.08}_{-1.85}$$\times$10$^{-10}$  &  884.21$^{+1037.99}_{-271.82}$  &  1.64$^{+0.01}_{-0.00}$  &  0.11$^{+0.02}_{-0.00}$  \\
 
 H$_{2}$CO (2$_{1,2}$ - 1$_{1,1}$) &  1.13$^{+1.21}_{-0.00}$$\times$10$^{-10}$  &  7.16$^{+7.36}_{-1.96}$$\times$10$^{-8}$  &  19876.12$^{+83.69}_{-10210.59}$  &  1.43$^{+0.03}_{-0.82}$ & 0.10$^{+0.02}_{-0.00}$   \\
 H$_{2}$CO (2$_{1,2}$ - 1$_{1,1}$) extended  &  3.41$^{+49.98}_{-0.35}$$\times$10$^{-14}$  &  7.87$^{+8.06}_{-0.41}$$\times$10$^{-9}$  &  5219.93$^{+1264.67}_{-2456.96}$  &  1.11$^{+0.55}_{-0.93}$ & 0.09$^{+0.05}_{-0.09}$   \\
 \hline
\label{modres}
\end{tabular}}

\tablecomments{The LOC + MCMC parameters and their errors are presented. For details on each of the models refer to the different subsections enclosed in Section \ref{results3}.}
\end{center}
\end{table*}

We have 5 parameters explored by the MCMC: $a_{in}$, $a_{out}$, $r$, $f_{v}$ and  $\sigma_{turb}$. The prior probability distributions are set as:

 \begin{itemize}
     \item 10$^{-15}$ $<$ $a_{in}$ $<$ 10$^{-6.5}$
     \item 10$^{-15}$ $<$ $a_{out}$ $<$ 10$^{-6.5}$
     \item 500 $<$ $r$ $<$ 20000 (au)
     \item 0 $<$ $f_{v}$ $<$ 2
     \item 0 $<$  $\sigma_{turb}$ $<$ 0.35 (km/s).
 \end{itemize}

These limits are set from previous knowledge of the physical, kinematic and chemical nature of the source. For example, scaling the velocity profile more than the limit specified above would be incompatible with previous observations and modelling of CO and H$_{2}$O towards L1544 \citep{keto:15}. We also impose the condition that $a_{in}$ cannot be larger than $a_{out}$ as we expect the molecules studied here to be depleted towards the centre of the core and, therefore, be less abundant in that area. The prior distribution is a uniform distribution, meaning that all values in the ranges mentioned above have the same prior probability.\ 

\newpage

\section{Results} \label{results3}

The modelling results for the observed molecular transitions are separated into two Sections: rarer isotopologue transitions and main isotopologue transitions (Sections \ref{thin} and \ref{thick}, respectively). The LOC + MCMC parameter values for the models described in the following sections are summarised in Table \ref{modres}. \

\subsection{Rarer Isotopologue Transitions}\label{thin}

Due to the intrinsically lower fractional abundances of the rare isotopologues, C$^{34}$S, $^{34}$SO and HC$^{17}$O$^{+}$ are expected to emit only from the core (i.e. within 0.32 pc). We do not expect their transitions to trace the extended structure, as traced with the HCO$^{+}$ $J$ = 1 - 0 transition \citep{redaelli:22b}. Thus, for the modelling of rarer isotopologue transitions presented in the next subsections, we use the non-extended model described in Section \ref{models}.

\subsubsection{C$^{34}$S}\label{c34s}

\begin{figure}[h]
\centering
\includegraphics[width=9cm]{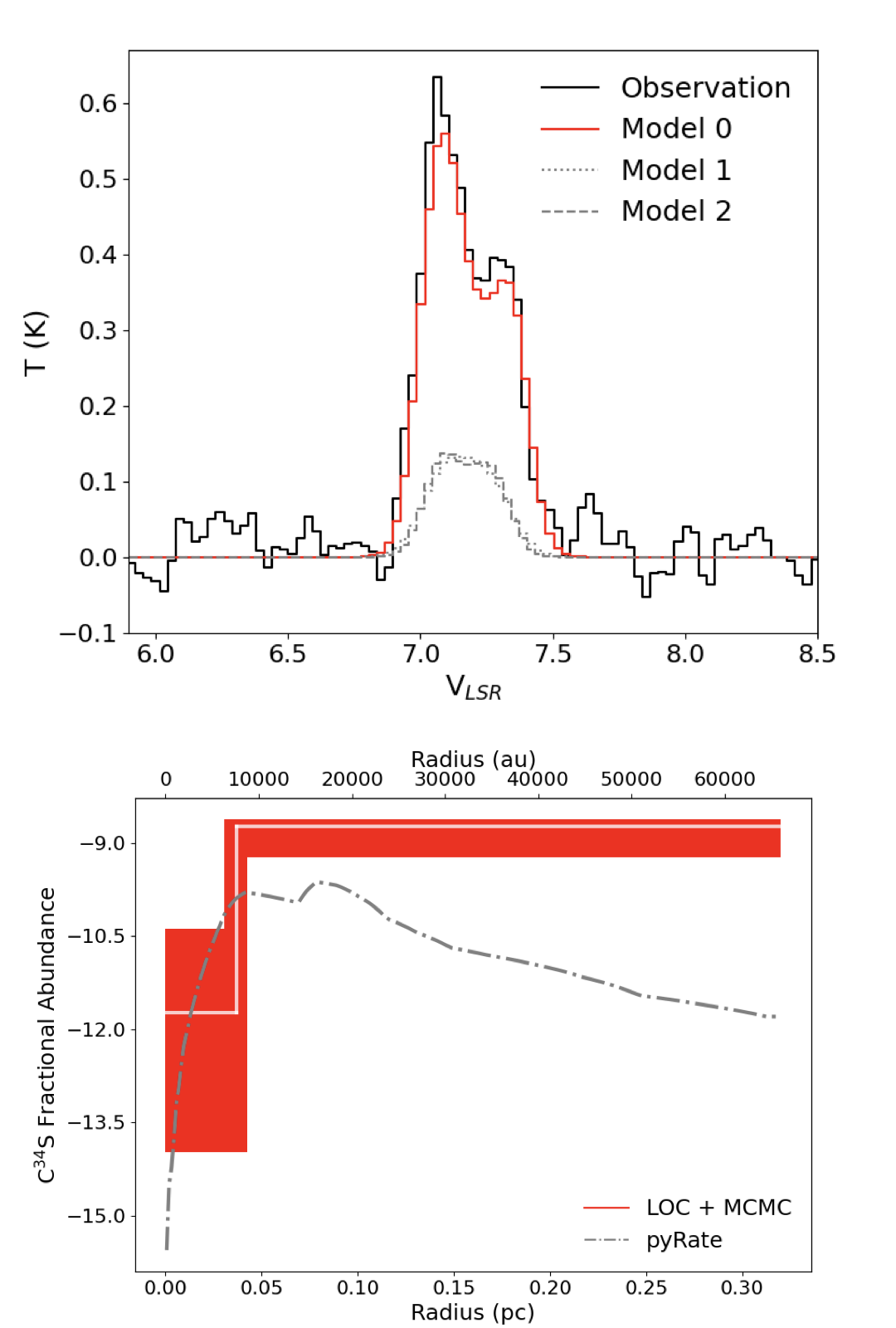}
\caption{\textit{Top panel:} Comparison of the C$^{34}$S (2 - 1) spectra computed with different models. The spectrum computed with the LOC + MCMC approach is plotted in a red solid line ( Model 0). The spectrum computed with the pyRate abundance profile and the resulting LOC + MCMC $f_{v}$ and  $\sigma_{turb}$ values is plotted with a dotted grey line ( Model 1). The C$^{34}$S spectrum computed with the pyRate fractional abundance profile and the pre-stellar core physical model velocity profile (\citealt{keto:15}) and  $\sigma_{turb}$ = 0.075 km/s, is plotted in a grey dashed line ( Model 2). Finally, the observations are plotted with a black solid line. \textit{Bottom panel:} C$^{34}$S fractional abundance profile obtained from the LOC + MCMC modelling of the observed line profile is plotted in a solid white line, over a red solid band, indicating the model uncertainties. This is compared with the fractional abundance profile obtained from pyRate plotted in a grey dashed-dotted line.}
\label{c34sspec}
\end{figure}

The C$^{34}$S (2 - 1) transition presents an asymmetric line profile indicative of infall (Figure \ref{obspap3}). To the best of our knowledge, this is the first time that the C$^{34}$S (2 - 1) line has been seen to show clear self absorption.  In \cite{tafalla:98}  this self absorption was hinted at (see Figure 3) but the S/N ratio was relatively low. Although the C$^{34}$S (2 - 1) line has a profile typical of an optically thick line tracing a contracting core, we are including this transition in this Section, as it is a rarer isotopologue and its optical depth ($\tau$ $\sim$  4.22) is however significantly lower than the CS (2 - 1) line ($\tau$ $\sim$  105.1). For further information refer to Appendix \ref{optdepth}. The LOC + MCMC approach reproduces well the observations (observations in black, model in red; top panel in Figure \ref{c34sspec}). The LOC + MCMC parameter values and their uncertainties are listed in Table \ref{modres}. The resulting corner plot is presented in Figure \ref{c34scpp}.\

The comparison of the fractional abundance profile used for the line profile from LOC + MCMC and the fractional abundance profile computed with pyRate is presented in the bottom panel of Figure \ref{c34sspec}. As pyRate does not take into account sulphur isotope chemistry, we scale down the CS fractional abundance profile by the [$^{32}$S/$^{34}$S] = 22 ratio \citep{wilson:94}.\ 

The radius where the C$^{34}$S abundance drops due to depletion in the centre of the core found with the LOC + MCMC modelling  is shifted to larger radii with respect to the radius computed with chemical models. The C$^{34}$S $a_{out}$ value found with the LOC + MCMC modelling is a factor $\sim$  9 larger than the highest C$^{34}$S  abundance from pyRate. The high abundances derived with the LOC + MCMC are connected to the high optical depth derived for this species.\ 

In order to see the effect of the abundance profile used on the output line profile, we compare the results of the LOC + MCMC modelling approach,  which we will from now on address as Model 0 (red solid line, top panel Figure \ref{c34sspec}), with the spectrum computed with the abundance profile from pyRate and the resulting LOC + MCMC $f_{v}$ and  $\sigma_{turb}$ values (Model 1; dotted grey line in the top panel of Figure \ref{c34sspec}). The Model 1 spectrum shows a  single-peaked line profile and under-reproduces the observed line intensity by a factor of $\sim$ 6 with respect to the observations.\

Moreover, to see the effect of the LOC + MCMC $f_{v}$ and  $\sigma_{turb}$ values on the modelled line profile with respect to the line profile computed by using the original velocity profile , from the physical core model described in \citealt{keto:15} and commonly assumed  $\sigma_{turb}$ = 0.075 km/s, we compute a third spectrum by using both the abundance profile from pyRate, the velocity profile from \citealt{keto:15}, and by setting  $\sigma_{turb}$ = 0.075 km/s (Model 2; dashed grey line in the top panel of Figure \ref{c34sspec}). This spectrum shows a line profile tending to a single-peak morphology and underestimates the line intensity with respect to the observations by a similar factor as Model 1.\

\subsubsection{$^{34}$SO}\label{34so}

The $^{34}$SO (2,3 - 1,2) transition has a single-peaked component with an asymmetric shoulder at $\sim$ 6.9 km/s (Figure \ref{obspap3}). Note that this line profile is not related to infall asymmetry. The infall asymmetry profile is seen as two-peaked line profiles with a less-intense red component compared to the blue component. Here, the asymmetric shoulder appears towards lower V$_{LSR}$ (blue part).  We looked for possible line candidates that fall close to the $^{34}$SO rest frequency without success. As of now the physical reason behind this asymmetric shoulder is not clear.The central velocity of this transition is shifted by 0.23 km/s with respect to the V$_{LSR}$ of the core when using the frequency value taken from the CDMS (97715.3170 $\pm$ 0.0500 MHz). Thus, the $^{34}$SO (2,3 - 1,2) transition was remeasured with the CAS Absorption Cell (CASAC) in the laboratory of the Center for Astrochemical Studies (CAS) in Garching bei M{\"u}nchen, Germany. The CASAC setup provides higher frequency accuracy measurements compared to the previously reported measurements for this molecule in the literature. The rest frequency was measured to be 97715.395 $\pm$ 0.023 MHz (Table \ref{obsdat}).\ 

\begin{figure}[h]
\centering
\includegraphics[width=9cm]{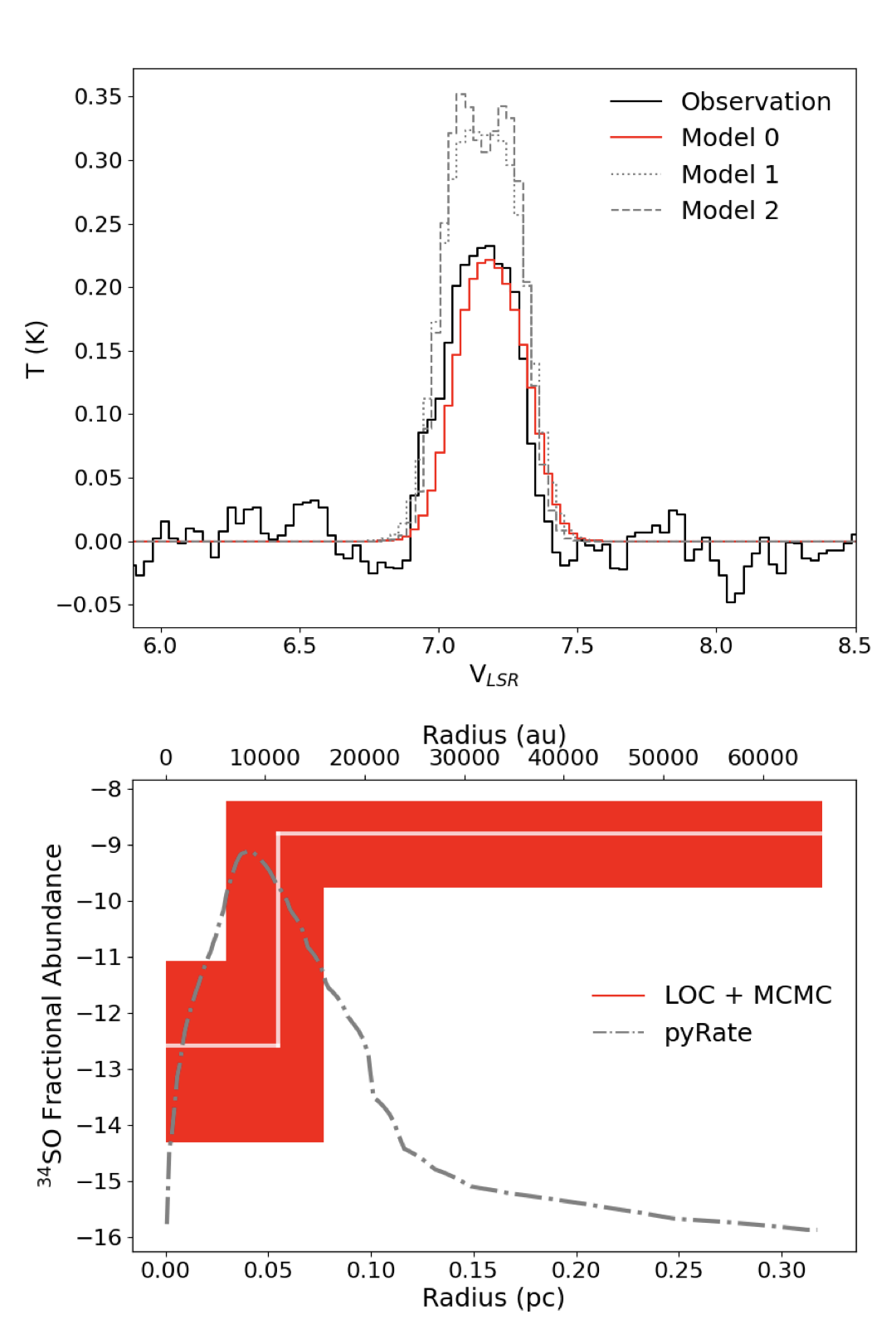}

\caption{Same as Fig. \ref{c34sspec}, but for the $^{34}$SO (2,3 - 1,2) transition.}
\label{34sospec}
\end{figure}

The resulting model fits the observed transition well, excluding the asymmetric shoulder (observations in black, Model 0 in red, top panel in Figure \ref{34sospec}). Nevertheless,  most of the variables are not constrained, making the corner plot deviate from Gaussian profiles to more flattened profiles (Figure \ref{34socpp}). The resulting Model 0 parameters are presented in Table \ref{modres}. The $a_{in}$ parameter was not constrained, appearing as a flat profile in the corner plot.  Moreover, the $r$, $f_{v}$ and  $\sigma_{turb}$ parameters do not show strong constraints. The $a_{out}$, on the other hand, appears better constrained with a visible peak in the corner plot. These uncertainties can also be seen represented with a broad red band in the bottom panel of Figure \ref{34sospec}. The asymmetric shoulder may be adding uncertainty to the modelled parameters, especially the ones directly linked with the line width. \

The comparison of the fractional abundance profile used for the LOC + MCMC line profile modelling and the fractional abundance profile computed with pyRate is presented in the bottom panel of Figure \ref{34sospec}.\ 

As mentioned above, the radius where the $^{34}$SO abundance drops due to depletion in the centre of the core found with the LOC + MCMC modelling presents large errors (seen as a broad red vertical band in the bottom panel of Figure \ref{34sospec}).   The depletion radius is larger in the LOC + MCMC model than in the pyRate abundance profile, even when taking into account the error bars. The $^{34}$SO $a_{out}$ value found with the LOC + MCMC modelling is a factor of $\sim$ 3 higher than the highest $^{34}$SO  abundance from pyRate. Nevertheless, the $a_{out}$ LOC + MCMC value and the pyRate abundance in this region agree within errors. \ 

As done for C$^{34}$S, we compare the spectra computed with  the different abundance profiles and physical parameter combinations, Model 0, 1 and 2 (for more information refer to Section \ref{c34s}).  Model 1 (dotted grey line in top panel, Figure \ref{34sospec}) overestimates the intensity with respect to the spectrum computed with  Model 0 by a factor of $\sim$ 1.4 (red solid line in top panel, Figure \ref{34sospec}).  Model 2 is double-peaked and overestimates the line intensity by a factor of $\sim$ 1.5 (dashed grey line in top panel, Figure \ref{34sospec}). \

As mentioned above, the spectra computed with the abundances predicted by pyRate, Models 1 and 2 (grey spectra in top panel, Figure \ref{34sospec}), are more intense than the one computed with the step abundance profile constrained by the LOC + MCMC approach; Model 0 (red spectrum in top panel, Figure \ref{34sospec}). This is due to the fact that the pyRate abundance profile is significantly below that derived by the LOC + MCMC (dash-dotted grey and white lines, respectively, in bottom panel, Figure \ref{34sospec}). The line intensity results from a combination of factors including the $n_{H_2}$ of the emitting region. The spectra computed with the abundances predicted by pyRate, shown in grey in top panel, Figure \ref{34sospec}, arise from a region closer to the centre of the core (see the peak of the abundance profile; dash-dotted grey line in bottom panel, Figure \ref{34sospec}), with respect to the LOC + MCMC computed spectrum (red in top panel, Figure \ref{34sospec}), which emits from larger radii (white line in bottom panel, Figure \ref{34sospec}).

\subsubsection{HC$^{17}$O$^{+}$}\label{hc17o3}

\begin{figure}[h]
\centering
\includegraphics[width=9cm]{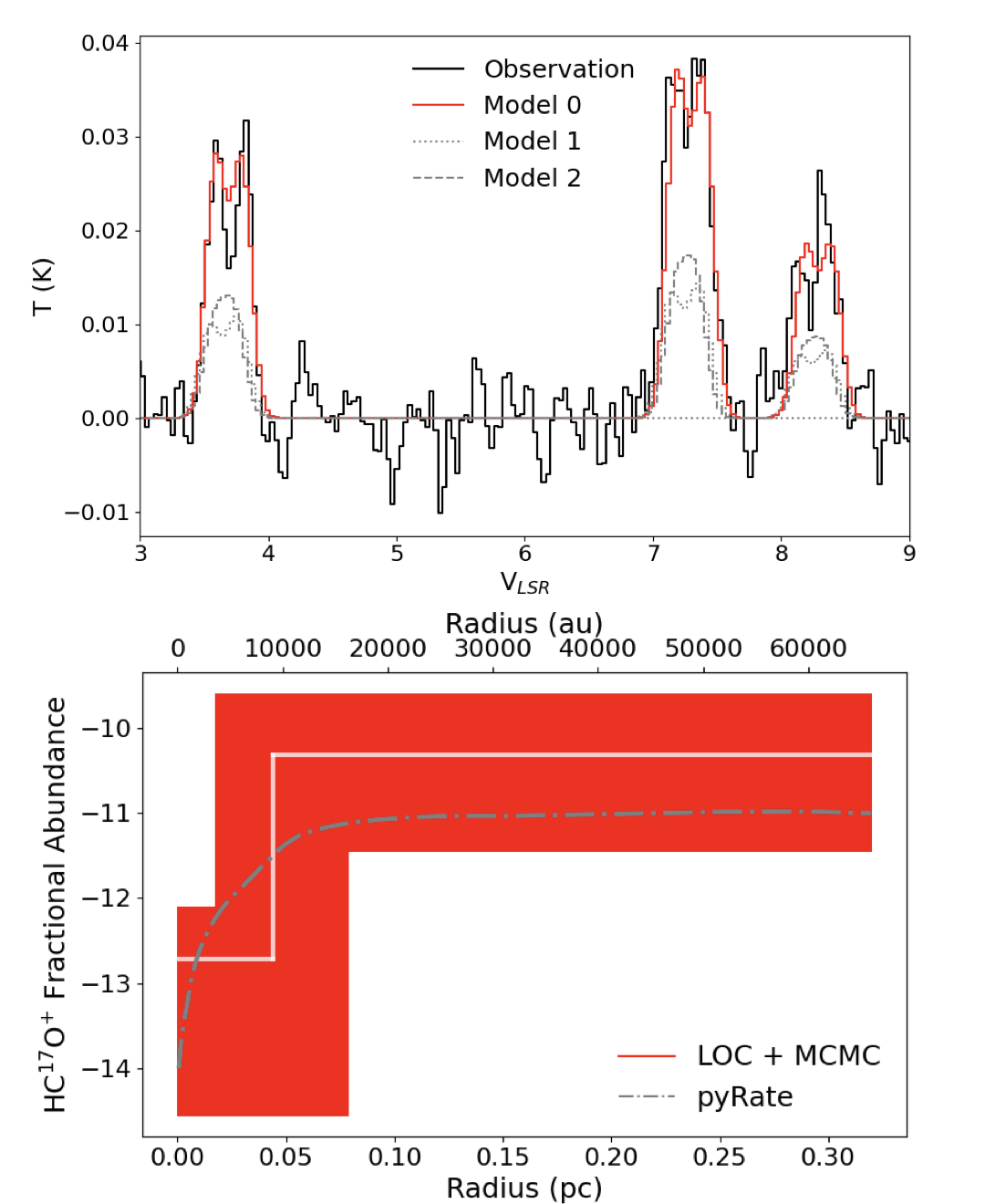}

\caption{Same as Fig. \ref{c34sspec}, but for the HC$^{17}$O$^{+}$  (1 - 0) transition. }
\label{hc17o+spec}
\end{figure}

The HC$^{17}$O$^{+}$ (1-0) transition, first presented in \cite{ferrerasensio:22}, is split into three hyperfine components ($F \rightarrow F^{'}$ 5/2 $\rightarrow$ 5/2, 5/2 $\rightarrow$ 7/2 and 5/2 $\rightarrow$ 3/2, from lower to higher velocity), which are in turn displaying a double-peaked line profile (Figure \ref{obspap3}).  The hyperfine structure of the HC$^{17}$O$^{+}$ (1-0) transition occurs because the nuclear spin of $^{17}$O (I = 5/2) and the molecular rotation are coupled, resulting in the splitting of the J = 1 rotational level. In \cite{ferrerasensio:22}, the observations were successfully reproduced with LOC using the pre-stellar core physical model described in \cite{keto:15} and the HC$^{17}$O$^{+}$ fractional abundance profiled scaled from the HCO$^{+}$ fractional abundance profile computed with pyRate. In this work, we test the LOC + MCMC approach using a step abundance profile (solid white line over a red solid band, Figure \ref{hc17o+spec}).\ 

The LOC + MCMC fits the line double-peaked profile slightly overestimating the intensity at the centre, not fully reproducing the depth of the dip, nevertheless, this discrepancy falls within the noise (Model 0, Figure \ref{hc17o+spec}). The variable values are presented in Table \ref{modres} and the corner plot in Figure \ref{hc17o+cpp}.\ 

The comparison of the LOC + MCMC fractional abundance profile and the fractional abundance profile computed with pyRate is presented in the bottom panel in Figure \ref{hc17o+spec}. As pyRate does not take into account oxygen isotope chemistry, we scaled down the HCO$^{+}$ fractional abundance profile by the [$^{16}$O/$^{17}$O] = 2044 ratio \citep{penzias:81, wilson:94}. The HCO$^{+}$ abundance profile used in this comparison is an updated abundance profile with respect to \cite{ferrerasensio:22}, as this better fits the HCO$^{+}$ observations towards L1544 \citep{redaelli:22b}.\ 

The radius where the HC$^{17}$O$^{+}$ abundance drops due to depletion in the centre of the core found with the LOC + MCMC modelling agrees with the radius at which the fractional abundance profile computed with chemical models presents a drop. Nevertheless, the abundance drop towards the core centre, as computed with the chemical models, is less steep compared to other molecules, and may not be well reproduced with the simplified step-abundance profile. The HC$^{17}$O$^{+}$ $a_{out}$ value found with the LOC + MCMC modelling is a factor $\sim$ 5 larger than the highest HC$^{17}$O$^{+}$ pyRate abundance. Nevertheless, the $a_{out}$ LOC + MCMC value and the pyRate abundance in this region agree within errors.  \

 A comparison between the simulated line profiles from the three models is shown in Figure \ref{hc17o+spec}. Model 1 (dotted grey line) underestimates the intensity by a factor of $\sim$ 3 and gives double-peaked hyperfine components.  Model 2 is also less intense by a similar factor, and the hyperfine components appear single-peaked (dashed grey line in the top panel in Figure \ref{hc17o+spec}). \

\subsection{Main Isotopologue Transitions}\label{thick}

The molecules included in the next subsections are abundant enough to be potentially present in the extended structure beyond 0.32 pc. Our approach is first to model the main isotopologue transitions arising from abundant molecules with the non-extended model and then test the extended model if the previous fails to reproduce the observations.\ 

\subsubsection{CS}\label{cs}
\begin{figure}[h]
\centering
\includegraphics[width=9cm]{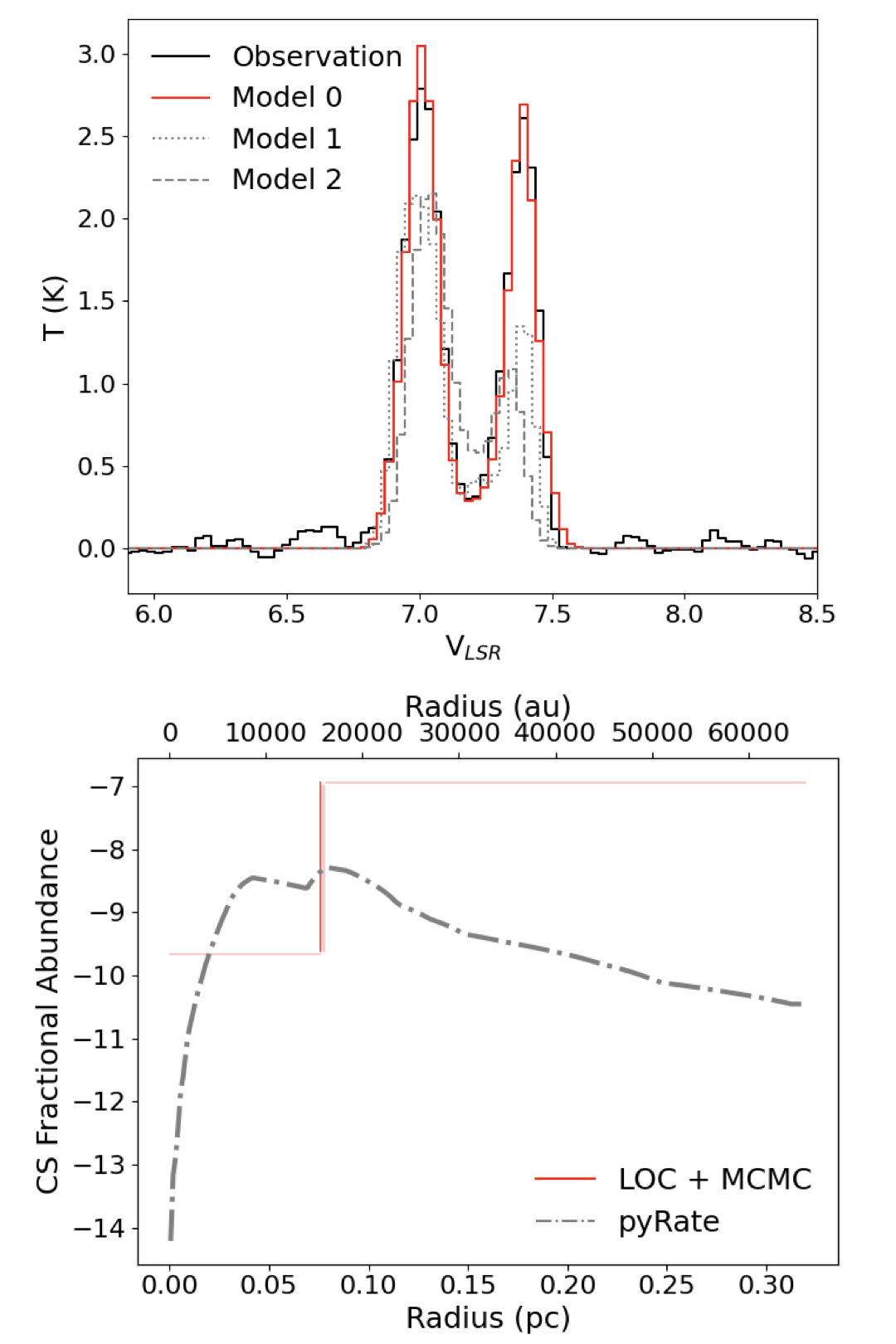}
\caption{Same as Fig. \ref{c34sspec}, but for the CS (2 - 1) transition non-extended model.}
\label{csp}
\end{figure}

The CS (2 - 1) transition is double-peaked with a dip that almost reaches the baseline (Figure \ref{obspap3}). This transition was originally presented in \citet{tafalla:98}. The blue and red peaks have similar intensities. This transition was fit with the non-extended model described in Section \ref{models}, and the results are presented in the red spectrum in the top panel in Figure \ref{csp}.

The model fits well the observed line profile (black spectrum in the top panel in Figure \ref{csp}).  Moreover, the variables are highly constrained compared to the ones found for the C$^{34}$S line (see corner plot in Figure \ref{cscpp}  and abundance profile on the bottom panel in Figure \ref{csp}). The resulting LOC + MCMC parameters are presented in Table \ref{modres}.  The $r$, $f_{v}$ and  $\sigma_{turb}$ parameters in particular present low errors, which translate to sharp columns in the corner plot (see Table \ref{modres}  and Figure \ref{cscpp}).  Moreover, the $f_{v}$ and  $\sigma_{turb}$ values were found to be 1.99 $^{+0.00}_{-0.01}$ and 0.01 $^{+0.00}_{-0.01}$, respectively  which are close to the boundaries (2 and 0) of the used prior distribution.\ 

The comparison of the fractional abundance profile used for the non-extended model and the fractional abundance profile computed with pyRate is presented in the bottom panel in Figure \ref{csp}.\ 

 The radius where the CS abundance drops due to depletion in the centre of the core found with the non-extended LOC + MCMC modelling appears shifted towards larger radii with respect to the chemical models. The CS $a_{out}$ value found with the LOC + MCMC modelling is a factor of $\sim$ 20 higher than the largest abundance from the chemical model.\ 

We compare the spectra computed with  Model 0 with Models 1 and 2 in Figure \ref{csp}. Model 1 (dotted grey line, top panel in Figure \ref{csp}) underestimates the line strength by a factor of  $\sim$ 1.4 but reproduces well the self-absorption dip.  Model 2 (dashed grey line, top panel in Figure \ref{csp}) also underproduces the line intensity by a factor of $\sim$ 1.4, but less than for the model using the resulting LOC + MCMC $f_{v}$ and  $\sigma_{turb}$ values.\

 As the non-extended model seems to be sufficient to reproduce the self-absorption dip, which is expected to be caused by the outermost layers of the core, we do not model CS (2 - 1) with the extended model. With the idea of better constraining the CS model parameters, we perform a combined modelling of the CS and C$^{34}$S (2 - 1) transitions. This test does not bring additional constraints with respect to the CS and C$^{34}$S separated models. For more information, see section \ref{csc34s} in the Appendix.\

\subsubsection{SO}\label{so}
\begin{figure}[h]
\centering
\includegraphics[width=9cm]{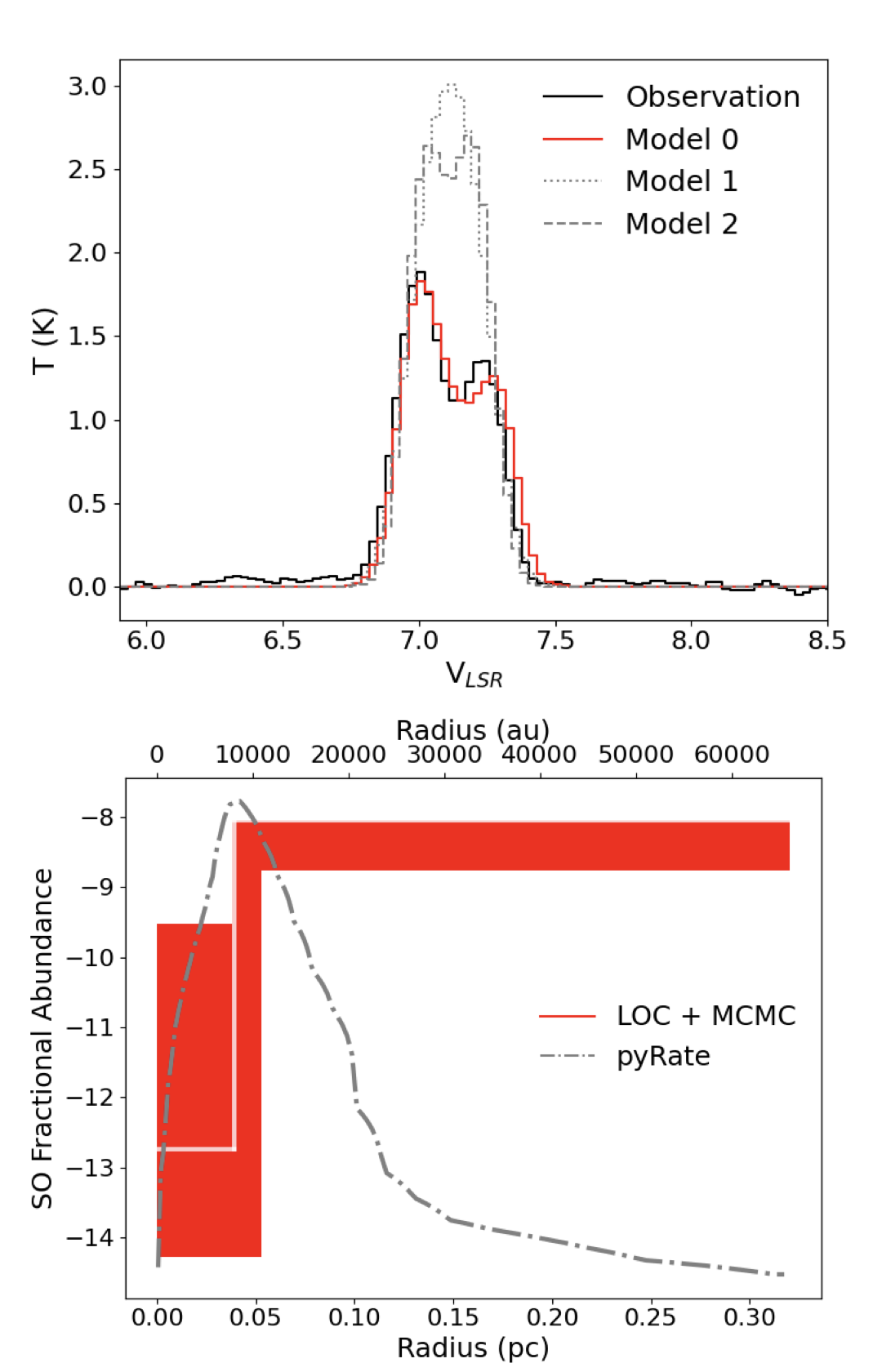}

\caption{Same as Fig. \ref{c34sspec}, but for the SO (2,3 - 1,2) transition.}
\label{sop}
\end{figure}

The SO (2,3 - 1,2) transition presents the characteristic infall asymmetry line profile (Figure \ref{obspap3}). Unlike other optically thick lines introduced in this work (CS, HCO$^{+}$ and H$_{2}$CO), SO presents a shallow self-absorption feature. The SO (2,3 - 1,2) transition is fitted with the non-extended model. As we can see in the top panel in Figure \ref{sop},  Model 0 (in red) fits well the observations (in black). The model parameters are presented in Table \ref{modres} and in the corner plot in Figure \ref{socpp}.\

The comparison of the fractional abundance profile used for the LOC + MCMC and the fractional abundance profile computed with pyRate are presented in the bottom panel in  Figure \ref{sop}.\ 

The radius where the SO abundance drops due to depletion in the centre of the core found with the LOC + MCMC modelling is shifted to larger radii with respect to the radius of the fractional abundance profile computed with chemical models. The $a_{out}$ SO value found with the LOC + MCMC modelling is a factor of $\sim$ 2 lower compared to the abundance of SO in the chemical model.\ 

The comparison of the spectra computed with  Model 0 with Models 1 and 2 is presented in the top panel in Figure \ref{sop}.  Model 1 results in a single-peaked line and overestimates the line intensity by a factor of $\sim$ 2.  Model 2 (dashed grey line in top panel in Figure \ref{sop}) also overestimates the line intensity by a factor of $\sim$ 1.7.\

Furthermore, as for the case of CS, we perform a combined fit of SO and $^{34}$SO to try to better constrain the parameters found with the individual models of the molecules.  This test does not add additional constraints (see Section \ref{so34so} in the Appendix).\ 

\subsubsection{HCO$^{+}$}\label{hco}

The HCO$^{+}$ (1 - 0) line shows a pronounced self-absorption dip that reaches the spectral baseline (third-column, first-row panel in Figure \ref{obspap3}). Moreover, there is a blue excess feature centred at $\sim$ 6.9 km/s. With the LOC RT model utilized here, we reproduce the HCO$^{+}$ (1-0) fit performed by \cite{redaelli:22b} using the RT code MOLLIE \citep{keto:05}. This is shown in section \ref{hco+red} of  the Appendix.\ 

In this Section, we test the LOC + MCMC methodology to fit the HCO$^{+}$ observations. As previous works pointed out the need for an external layer \citep{redaelli:19a, redaelli:22b}, we modelled HCO$^{+}$ with the extended model described in Section \ref{models} with some adjustments. The velocity and $n_{H_{2}}$ were fixed to -0.05 km/s and 27 cm$^{-3}$, respectively \citep{redaelli:22b}. The abundance of HCO$^{+}$ in the external layer is kept constant with the value of the abundance profile predicted by pyRate at 0.32 pc, as done in \cite{redaelli:22b}.\ 

\begin{figure}[h]
\centering
\includegraphics[width=9cm]{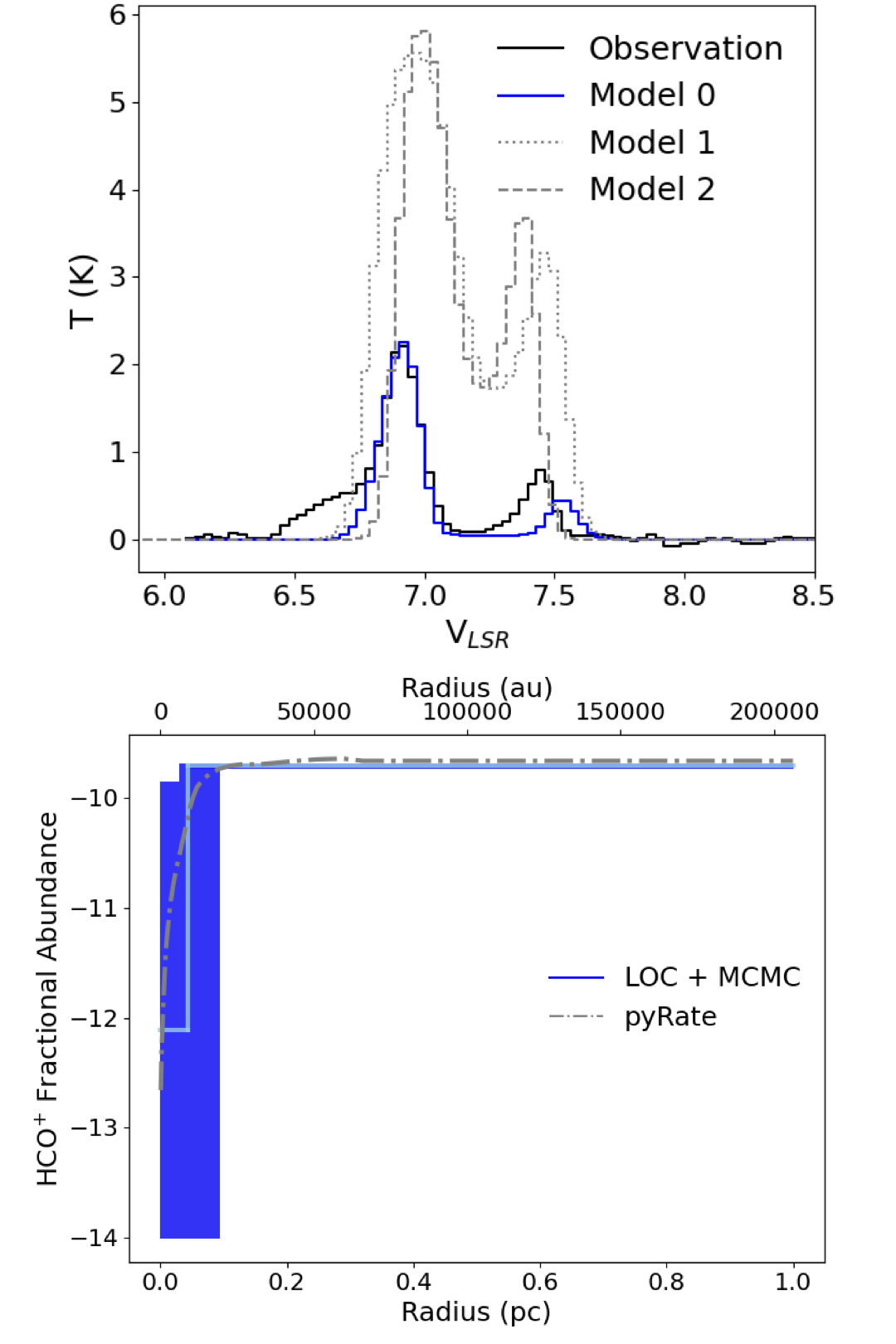}

\caption{Same as Fig. \ref{c34sspec}, but for the HCO$^{+}$ (1 - 0) transition. Notice that in this case, as the extended model was used, both the Model 0 spectrum as well as the abundance profile uncertainty bands are plotted in blue. For better contrast, the white line showing the abundance profile is plotted in this case with a sky blue colour.}
\label{hco+p}
\end{figure}

This approach reproduces the HCO$^{+}$ (1 - 0) fairly well, except for the separation between the two peaks, which is overestimated by the model (top panel, Figure \ref{hco+p}). The model parameters are presented in Table \ref{modres} as well as in the corner plot in Figure \ref{hco+cpp}.  The parameters are fairly well constrained with the exception of $a_{in}$, as expected from an optically thick transition, which does not trace the core inner regions. \ 

 The comparison of the fractional abundance profile used for the LOC + MCMC and the fractional abundance profile computed with pyRate are presented in the bottom panel in Figure \ref{hco+p}.  The $r$ value agrees within errors with the pyRate abundance profile drop. The $a_{out}$ parameter found is of the same order of magnitude as the pyRate one.\ 

In the top panel of Figure \ref{hco+p} the spectra computed with Models 0, 1 and 2 are presented. Both Models 1 and 2 overestimate the line intensity by a factor of $\sim$ 3 and do not reproduce the depth of the absorption feature. Moreover, Model 1 results in a broader line profile than the one observed.\

In addition, we also tested fitting HCO$^{+}$ and HC$^{17}$O$^{+}$ together. This test does not add additional information than the one found for the single HCO$^{+}$ and H$^{17}$CO$^{+}$ model fits (see Section \ref{hco+hc17o+} in  the Appendix).\

\subsubsection{\ce{H2CO}}\label{h2co}

\begin{figure}[H]
\centering
\includegraphics[width=9cm]{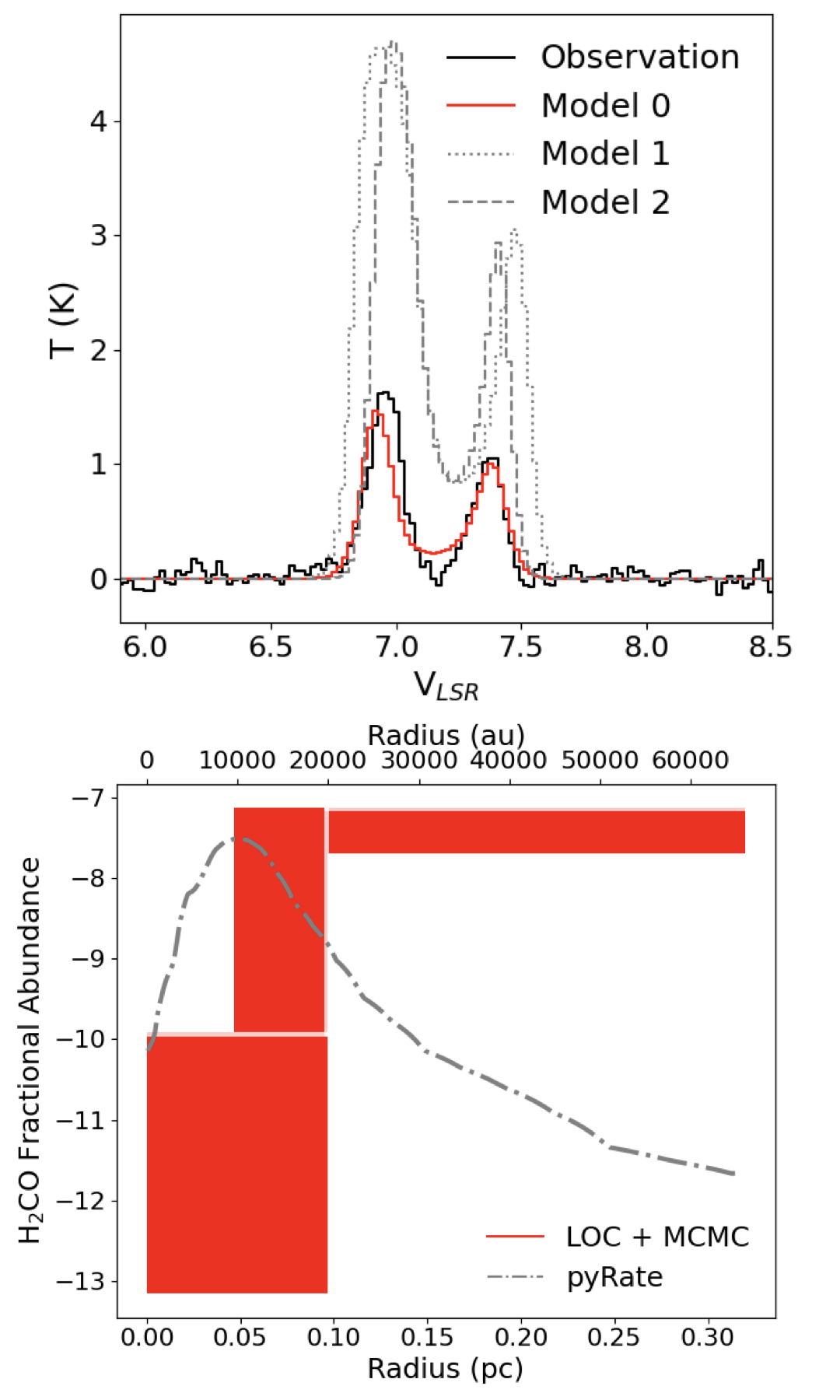}

\caption{Same as for Fig. \ref{c34sspec}, but for the H$_{2}$CO (2$_{1,2}$ - 1$_{1,1}$) transition.}
\label{h2cop}
\end{figure}

The H$_{2}$CO  (2$_{1,2}$ - 1$_{1,1}$) transition presents a self-absorption dip that reaches the baseline (fourth-column, first-row panel in Figure \ref{obspap3}). Moreover, the blue component is brighter than the red component. The asymmetry of the blue and red peaks indicates that H$_{2}$CO traces contracting gas. The results for the non-extended modelling are presented in red in Figure \ref{h2cop}.   The LOC + MCMC modelling reproduces well the observations. The corner plot presents for all of the parameters, except for $a_{in}$, two peaks indicating two possible solutions (Figure \ref{h2cocpp}). The model parameter values are listed in Table \ref{modres}.\ 

 The comparison of the fractional abundance profile used for the LOC + MCMC and the fractional abundance profile computed with pyRate are presented in the bottom panel in Figure \ref{h2cop}.   The $r$ value is shifted to larger radii with respect to the drop in abundance in the pyRate abundance profile. The $a_{out}$ parameter value found agrees within errors with the highest value of the pyRate abundance profile.\

 As done with the other molecules, we compare the spectra computed with Model 0 with Models 1 and 2 (dotted grey line in top panel in Figure \ref{h2cop}).  Both Models 1 and 2 overestimate the line intensity by a factor of $\sim$ 2.5, do not reproduce the depth of the absorption feature and have broader line profiles than the observed one. \

\begin{figure}[H]
\centering
\includegraphics[width=9cm]{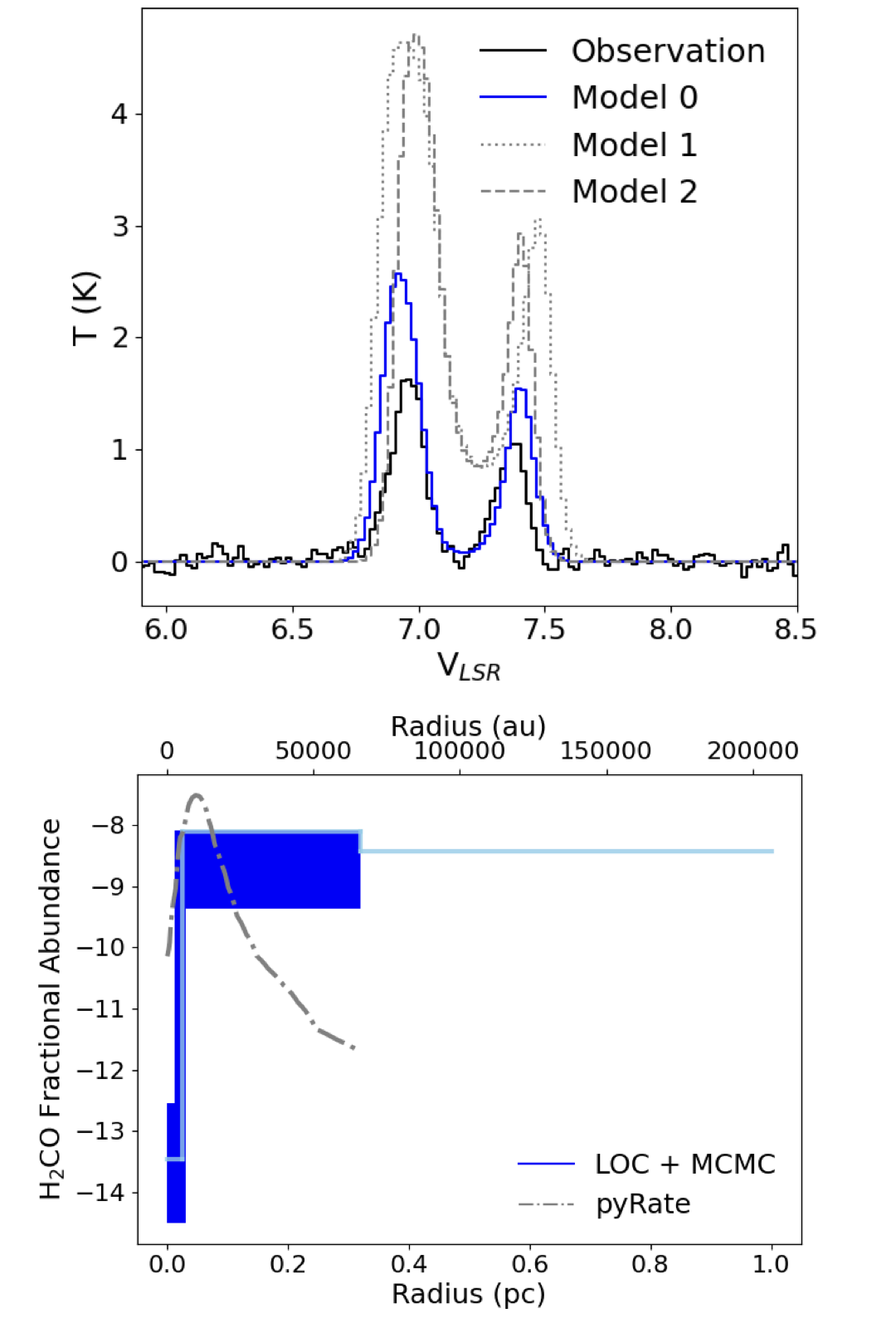}

\caption{Same as for Fig. \ref{c34sspec}, but for the H$_{2}$CO (2$_{1,2}$ - 1$_{1,1}$) transition. Notice that in this case the extended model is presented, and thus, Model 0 as well as the abundance profile uncertainties are plotted in blue. In the same way as for Fig. \ref{hco+p}, the abundance line profile computed with the LOC + MCMC is plotted with a sky blue colour for better contrast.  }
\label{h2coep}
\end{figure}

In view of these results, we test the extended model (see Figure \ref{h2coep}). The constant H$_{2}$CO abundance profile in the external layer is fixed to the value 3.7$\times$10$^{-9}$ as found for diffuse clouds \citep{snow:06}  The extended model  over-predicts the line intensity and the separation of the two peaks (blue dash-dotted line, top panel, Figure \ref{h2coep}). Nevertheless, this model presents an improvement with respect to the non-extended model when it comes to reproducing the self-absorption dip. This result implies that H$_{2}$CO needs to be present in the diffuse surrounding cloud to reproduce the depth of the self-absorption feature, as seen for HCO$^{+}$ (1 - 0) \citep{redaelli:22b} The corner plot for the extended model (Figure \ref{h2coecpp}) shows  a larger scatter  \

 The comparison of the fractional abundance profile used for the LOC + MCMC and the fractional abundance profile computed with pyRate are presented in the bottom panel in Figure \ref{h2coep}.   The pyRate abundance profile extends to 0.32 pc only. Moreover, as the $a_{ext}$ is a fixed value only, a sky blue line is plotted without uncertainties. The $r$ value agrees within error with the radii of the pyRate abundance profile. The $a_{out}$ parameter value is a factor of 4 lower than the highest value of the pyRate abundance profile.\

 The resulting spectra computed with Models 1 and 2 are presented in Figure \ref{h2coep}).  Both Models 1 and 2 overestimate the line intensity by a factor of $\sim$ 1.8 and do not reproduce the depth of the absorption feature and have broader line profiles than observed \

\subsubsection{c-C$_{3}$H$_{2}$}\label{cc3h2}

The c-C$_{3}$H$_{2}$ (2$_{1,2}$ - 1$_{0,1}$) transition also presents a self-absorption feature but not as pronounced as for CS, HCO$^{+}$ and H$_{2}$CO (fourth-column, second-row in Figure \ref{obspap3}). The blue and red peaks show similar intensities, hinting at this transition arising from a static layer, which was previously deduced from the same transition in \cite{tafalla:98}. This is also seen for CS (2 - 1) (Section \ref{cs}).\ 

\begin{figure}[H]
\centering
\includegraphics[width=9cm]{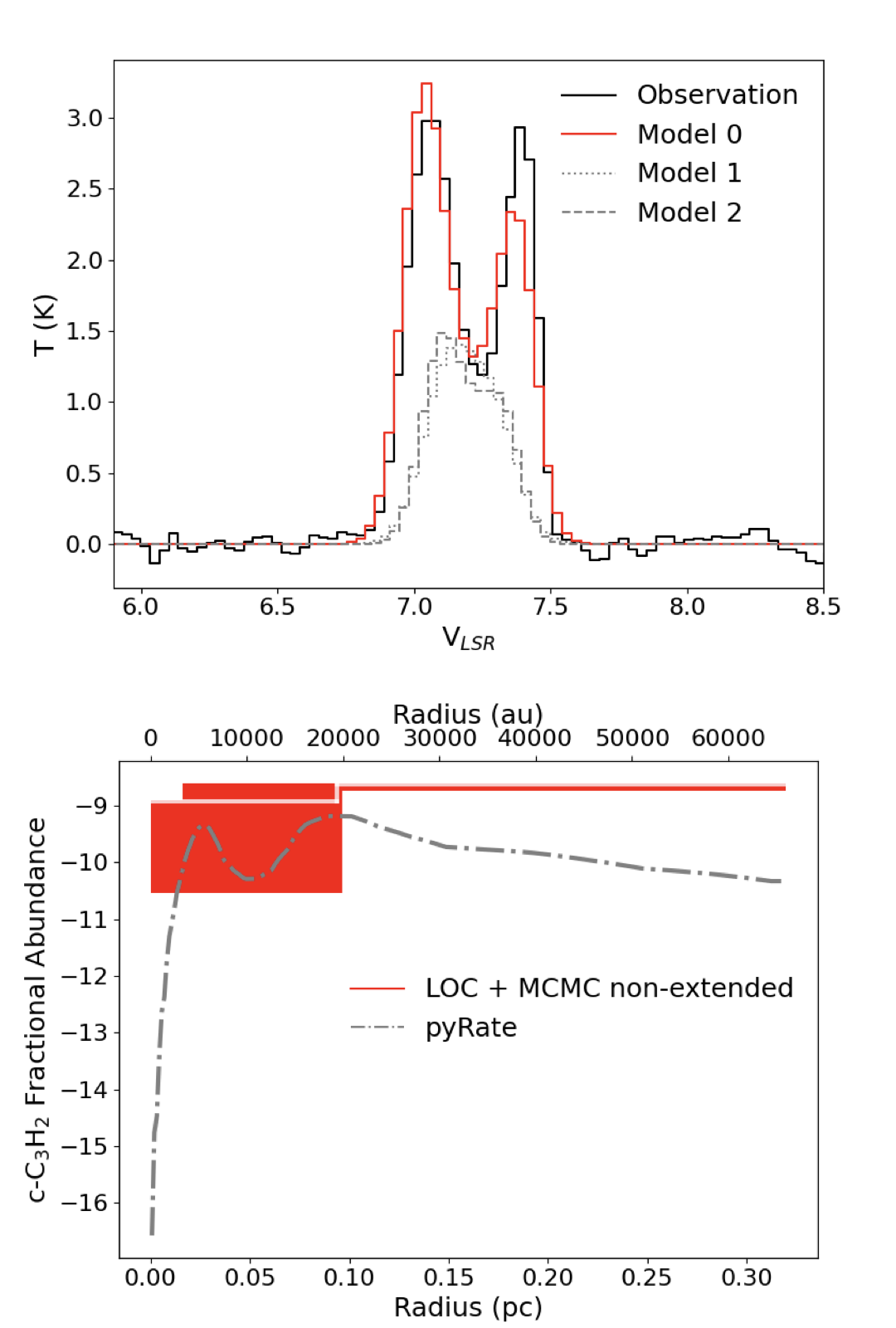}

\caption{Same as for Fig. \ref{c34sspec}, but for the c-C$_{3}$H$_{2}$ (2$_{1,2}$ - 1$_{0,1}$) transition.}
\label{cc3h2p}
\end{figure}

The results of the non-extended model are shown with a solid red line in the top panel in Figure \ref{cc3h2p}. The model reproduces well the intensity of the blue peak but underestimates the intensity of the red peak by a factor of $\sim$ 1.5. The parameters are presented in Table \ref{modres} as well as in the corner plot in Figure \ref{cc3h2cpp}.\ 

 The comparison of the fractional abundance profile used for the LOC + MCMC and the fractional abundance profile computed with pyRate are presented in the bottom panel in Figure \ref{cc3h2p}.   The $r$ value is shifted to larger radii with respect to the main drop in abundance in the pyRate abundance profile. The $a_{out}$ parameter value found is $\sim$ 3 times higher than the maximum value of the pyRate abundance profile.\

 The comparison of the spectra computed with Model 0 with Models 1 and 2 can be found in the top panel in Figure \ref{cc3h2p}.  Both Models 1 and 2 underestimate the line intensity by a factor of $\sim$ 2. Moreover, in Models 1 and 2 the double-peaked line profile is only hinted.\

\begin{figure}[H]
\centering
\includegraphics[width=9cm]{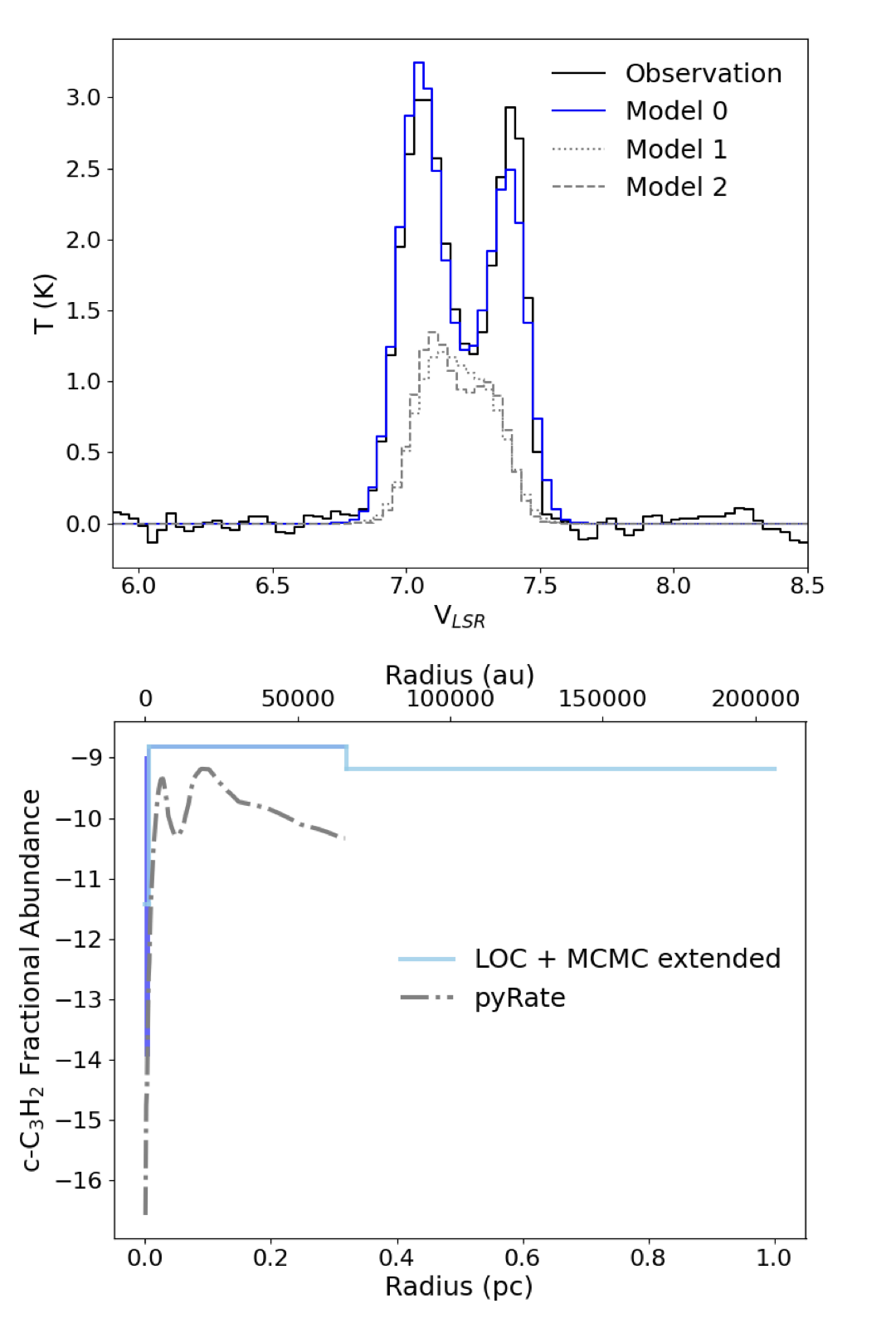}

\caption{Same as for Fig. \ref{c34sspec}, but for the c-C$_{3}$H$_{2}$ (2$_{1,2}$ - 1$_{0,1}$) transition. Notice that in this case the extended model is presented, and thus, Model 0 as well as the abundance profile uncertainties are plotted in blue. In the same way as for Fig. \ref{hco+p}, the abundance line profile computed with the LOC + MCMC is plotted with a sky blue colour for better contrast.}
\label{cc3h2ep}
\end{figure}

We also tested the extended model to see if this would improve the fit with respect to the non-extended model (see Figure \ref{cc3h2ep}). We set the c-C$_{3}$H$_{2}$ constant abundance value in the external layer to 6.4$\times$10$^{-10}$, as seen in diffuse clouds \citep{snow:06}.  The fits of the extended and non-extended models are similar (see Figure \ref{cc3h2p}). The parameters are presented alongside the corner plot in Figure \ref{cc3h2ecpp}.  With the exception of the $a_{in}$ and $r$, the parameters found for the non-extended and extended models are similar.\ 

 The comparison of the fractional abundance profile used for the LOC + MCMC and the fractional abundance profile computed with pyRate are presented in the bottom panel in Figure \ref{cc3h2ep}.   The $r$ value partially agrees with the drop in the pyRate abundance profile. The $a_{out}$ parameter value found is $\sim$ 2 higher than the maximum value of the pyRate abundance profile.\

  Both Models 1 and 2 underestimate the line intensity by a factor of $\sim$ 2. In this case the double-peaked line profile of Model 2 becomes more apparent.\

The c-C$_{3}$H$_{2}$ (2$_{1,2}$ - 1$_{0,1}$) transition is not constrained with the LOC + MCMC modelling approach. Additional transitions are needed to improve the fit and decrease the uncertainties in the model parameters. \

\newpage

\section{Discussion} \label{discussion3}

 In general the LOC + MCMC modelling is successful in reproducing the observed spectra. Nevertheless, not all of the parameters are constrained. Two types of physical models are tested in this manuscript; the extended and non-extended models. The non-extended model was used for all of the transitions except for the HCO$^{+}$ (1 - 0) one, which was directly fitted with the extended model as previous works showed the need to take into account an extended structure to fit the observed line profile \citep{redaelli:22b}.  All of the observed spectra are reasonably well fitted with the non-extended model. As a test, H$_{2}$CO (2$_{1,2}$ - 1$_{1,1}$) and c-C$_{3}$H$_{2}$ (2$_{1,2}$ - 1$_{0,1}$) were also modelled with the non-extended model. These lines were selected because their line profiles show self-absorption, suggesting they may trace an extended structure, and because their fits with the non-extended model showed room for improvement. The extended model fits improve the results from the non-extended model for the H$_{2}$CO (2$_{1,2}$ - 1$_{1,1}$) case. Thus, both H$_{2}$CO (2$_{1,2}$ - 1$_{1,1}$) and HCO$^{+}$ (1 - 0) seem to be tracing material outside 0.32 pc, commonly considered in radiative transfer modelling of L1544 as the edge of the cloud. Additionally, with the aim of constraining the parameters explored by the MCMC by giving more initial constraints, another approach was tested. This additional test consisted in fitting the main and rare isotopologues together. This does not improve the constraints on the modelled fractional abundance profiles (see Appendix).\ 

 The fact that the models fit well the observations visually but the parameters explored with the MCMC are not constrained indicates that there may be other possible parameter combinations that result in a good fit of the spectra. Thus, high-resolution spectra of single lines do not provide enough information to converge to one unique solution. A way to improve the constrain on the parameters would be to fit multiple lines for each species and isotopologue as done in \cite{redaelli:19a} and \cite{spezzano:25}.  Future work should aim to include multiple lines of each species, to provide more constraints for the fitting (see, e.g., \citealt{lin:22, jensen:24}).   \ 

\subsection{Evaluation of Rarer Isotopologue Fits}

 From the rarer isotopologue lines presented in Section \ref{thin}  the one with the most constrained $a_{in}$ is HC$^{17}$O$^{+}$ (1-0). This, alongside the higher value of $f_{v}$ found, which is interpreted as an enhanced gas contraction velocity towards the core centre (\citealt{ferrerasensio:22}),  could indicate that HC$^{17}$O$^{+}$ (1-0) is the line tracing the innermost part of the core out of the three rarer isotopologues, and is possibly the most optically thin. Another interesting result from the HC$^{17}$O$^{+}$ (1-0) modelling is the highly unconstrained $r$ parameter. The inability for the model to constrain $r$ may come from the abundance profile decreasing in a more gradual fashion as predicted by pyRate (dash-dotted profile on the lower panel of Figure \ref{hc17o+spec}).  This may also be the reason for the unconstrained $a_{out}$, which depends on $r$.\

 The unconstrained $a_{in}$ of C$^{34}$S (2 - 1) seems to indicate this line may be slightly optically thick, as traced by the blue asymmetric line profile observed, and not tracing the innermost part of the core as seen for HC$^{17}$O$^{+}$ (1-0). An additional test to determine the optical depth of this transition (see Appendi \ref{optdepth})  returned a value of $\tau$ $\sim$ 4.22, which is unexpectedly high. This is most likely due to the $a_{out}$ found for C$^{34}$S which is almost an order of magnitude higher than the peak abundance computed by pyRate (see dash-dotted grey line in the bottom panel of Figure \ref{c34sspec}).\

 The most striking result from the $^{34}$SO (2,3 - 1,2) LOC + MCMC modelling is the significant displacement of $r$ with respect to the abundance profile computed with pyRate (see dash-dotted grey line in the bottom panel of Figure \ref{34sospec}). Specifically, the radius where $^{34}$SO depletes computed with the LOC + MCMC lays at larger radii than the one shown in the fractional molecular abundance profile computed with chemical modelling. One has to keep in mind that $r$ was not well-constrained, but these results could also be pointing at a problem with the chemical network used to compute the SO, and consequently $^{34}$SO, fractional abundance profile.\  

 From these results, C$^{34}$S and $^{34}$SO seem to trace a region at similar radii in the inner part of L1544, while HC$^{17}$O$^{+}$ traces deeper regions towards the centre of the core. These transitions provide constraints on the molecular fractional abundance profiles and on the kinematic structure of the inner parts of L1544.\ 

\subsection{Evaluation of Main Isotopologue Fits}

 From the main isotopologue transitions, CS (2 - 1) is the one with the most constrained parameters. Nevertheless, upon inspection of the parameter values found by the LOC + MCMC modelling, the physical correctness of this solution seems doubtful. First of all, the values for $f_{v}$ and  $\sigma_{turb}$, 1.99 and 0.00 respectively, match the bounds set in the prior distribution of 2 and 0 respectively. This seems to indicate the assumed model is too simplistic for a convergence to be found. Moreover, as seen for C$^{34}$S, the value of $a_{out}$ is 2 orders of magnitude higher than the maximum abundance computed with pyRate (see dash-dotted grey line on the bottom panel of Figure \ref{csp}).  This seems to be the reason why the calculated CS optical depth is unexpectedly high ($\tau$ $\sim$ 105.1, see Appendix \ref{optdepth}).\

  As seen for $^{34}$SO, the calculated $r$ by the LOC + MCMC approach of SO is also significantly shifted towards larger radii (see dash-dotted grey line in the bottom panel of Figure \ref{sop}).\

Observationally, SO has an extended morphology in some pre-stellar cores (e.g. in L183, \citealt{swade:89, dickens:00} and \citealt{pagani:05}) and it could also be the case for L1544 (see Figure A.3 in \citealt{spezzano:17}). The drop in the outer parts of the core in the SO fractional abundance profile computed with PyRate, shown in a grey dashed line in the bottom panel in Figure \ref{sop}, can result from a combination of factors: photodissociation, "depleted" initial abundance assumption and uncertainties in sulphur chemistry. The "depleted" initial abundance assumption involves reducing the initial S abundance to avoid an excess of sulphur-bearing species in the inner part of the core to reproduce the observed sulphur depletion known as "the sulphur depletion problem" (e.g. \citealt{ruffle:99}). This results in an underestimation of the sulphur-bearing abundances on the outer parts of the core. At the edge of the core, where the visual extinction is  estimated to be of the order of 1 mag, ~61\% of SO destruction is mostly due to reactions such as SO + C$^{+}$, with ~38\% of the SO destruction resulting from photodissociation. In regions with Av = 2 mag most of the destruction comes from SO + C$^{+}$ and only ~4\% of the destruction is due to photodissociation. This seems to point out at a non-negligible effect of the initial S depletion abundance as well as possibly the incompleteness of the sulphur chemical network on the computed low SO abundance at the edge of the core. \

 The HCO$^{+}$ (1 - 0) transition is not that well reproduced compared to the other main isotopologue transitions. The complicated line profile seems to be affecting the line width of Model 0.\

 The H$_{2}$CO (2$_{1,2}$ - 1$_{1,1}$) transition is fairly well fitted by Model 0. Nevertheless, the parameter exploration by the MCMC resulted into a double peaked distribution indicating that there are two equally possible solutions to fit this line profile. In view of these results we tested the extended model, which resulted in a better reproduction of the self-absorption feature. Nevertheless, the constraint of the parameters was not improved.\

 The c-C$_{3}$H$_{2}$ (2$_{1,2}$ - 1$_{0,1}$) transition is fairly well fitted by Model 0 except for the red peak whose intensity is under-reproduced. As done for H$_{2}$CO we have tested the extended model to see if the fit would be improved, but in this case this resulted in a similar fit than the non-extended model.\

\begin{figure}[H]
\centering
\includegraphics[width=9cm]{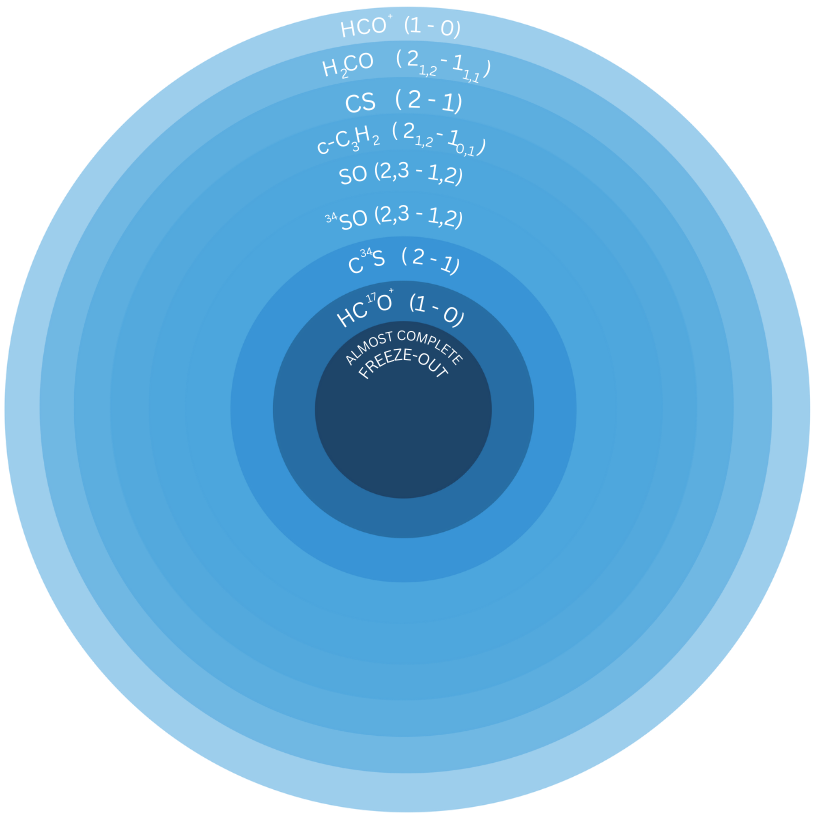}
\caption{Schematics of the L1544 layers traced by the different molecular transitions presented in the previous sections.  }
\label{onion}
\end{figure}

\subsection{Spatial Distribution of Molecular Emission in L1544}

All of the transitions presented in this manuscript trace different parts of the L1544 core, which have different physical, chemical and kinematic properties. To illustrate this point, we sketch in Figure \ref{onion} the relative location of the studied molecular transitions within L1544. The structure in Figure \ref{onion} results from the interpretation of the observed line profiles. As not all of the LOC + MCMCs' radii were well constrained, exact spatial distributions of the molecules cannot be determined. That is also the reason why precise modified gas velocity and turbulence profiles to fit all of the transitions together cannot be constructed. The main factors which decide the position of the layers are the presence or absence of self absorption and its magnitude and the double peaked line profile tracing contraction. The presence of a self-absorbed feature requires a foreground layer that absorbs the radiation emitted on a background layer. Thus, molecular transitions that show a self absorbed line profile trace spatially more extended regions than molecular transitions that do not show self absorption, being then present in external layers. The more material in the foreground layer, the more emission can be absorbed resulting in a deeper self-absorption feature.\

 In the outermost layer we are placing HCO$^{+}$ (1 - 0), as it is the only transition whose model requires the molecule to be present beyond the commonly assumed edge of the cloud (0.32 pc). The HCO$^{+}$ (1 - 0) line profile is characterised by a very deep self-absorption feature that reaches the baseline level, as well as a pronounced blue line asymmetry. This indicates that the molecule is abundant in a foreground layer capable of absorbing background emission and that this layer is undergoing contraction. The second transition with a comparably deep self-absorption feature is H$_{2}$CO (2$_{1,2}$ - 1$_{1,1}$). The self-absorption in this transition also reaches the baseline, and like HCO$^{+}$, the self-absorption dip is just fitted by including material beyond the core's boundary. Thus, it is interpreted as tracing material beyond the dense core boundary. The blue asymmetry in its line profile similarly indicates contraction in the layer it traces. Following the self-absorption dip trend, the next transition is CS (2 - 1), which shows a deep but shallower self-absorption feature that does not reach the baseline. The comparable brightness of the blue and red peaks suggests that CS is tracing a static layer in the outer parts of the core.\

 The CS layer is followed by that of c-C$_{3}$H$_{2}$ (2$_{1,2}$ - 1$_{0,1}$), which also shows a self-absorption dip and similar blue and red peak brightness, indicating that this molecule is tracing a static middle-to-outer layer of the core. Next is SO (2,3 - 1,2), which displays the prototypical blue asymmetry line profile, suggesting contraction. However, the relatively weak self-absorption dip implies that this line originates from an intermediate region within the L1544 core. In addition, SO shows a significant shift in its depletion radius ($r$) compared to the predictions from chemical modelling, possibly pointing to incomplete sulphur chemistry or underestimated abundances in the outer layers.\

 The inner layers of the core are traced by the rarer isotopologues. Continuing inward, we find $^{34}$SO (2,3 - 1,2), which is the only single-peaked line in this study, suggesting that it is not abundant enough in the outer layers to produce self-absorption, nor is it tracing the very innermost, high-velocity region. It is therefore assigned to an intermediate position in the diagram. C$^{34}$S (2 - 1) and HC$^{17}$O$^{+}$ (1 - 0) probe the deepest layers of L1544. Both lines show double-peaked profiles that suggest they trace the contracting central regions, where freeze out is almost complete  \citep{caselli:22}. Among these, HC$^{17}$O$^{+}$ shows the highest velocity profile scaling ($f_v$), indicating it likely originates from the very centre of the core. It is thus placed in the innermost layer of the diagram, followed by C$^{34}$S at slightly larger radii.\

The comparison between the spectra computed with the LOC + MCMC approach, with the original fractional abundance profiles computed with pyRate and the resulting LOC + MCMC $f_{v}$ and  $\sigma_{turb}$ values, and the original fractional abundance and velocity profiles and  $\sigma_{turb}$ = 0.075 km/s, showed the need of modifying the abundance profiles and physical parameters used for the modelling of the high-sensitivity and high-spectral resolution observed lines towards L1544.\

\section{Conclusions}\label{conclusions3}

This work has presented new high-sensitivity and high-spectral resolution observations of the \ce{HCO+} (J = 1 - 0), CS (J = 2 - 1), C$^{34}$S (J = 2 - 1), \ce{H2CO} (J$_{K_{a},K_{c}}$ = 2$_{1,2}$ - 1$_{1,1}$), c-C$_{3}$H$_{2}$ (J$_{K_{a},K_{c}}$ = 2$_{1,2}$ - 1$_{0,1}$), SO (N,J = 2,3 - 1,2) and $^{34}$SO (N,J = 2,3 - 1,2) as well as HC$^{17}$O$^{+}$ (J = 1 - 0) from \citet{ferrerasensio:22} rotational transitions towards the dust peak of the L1544 pre-stellar core.\ 

 The modelling successfully reproduces most line profiles, but not all parameter distributions were well constrained by the MCMC, highlighting degeneracies in the modelling approach. The only two transitions requiring an extended envelope beyond 0.32 pc for a good fit are HCO$^{+}$ (1–0), consistent with previous works, and H$_{2}$CO (2$_{1,2}$ - 1$_{1,1}$). c-C${3}$H$_{2}$ showed self-absorption but did not benefit from an extended model. The combined modelling of the main and rare isotopologues did not improve parameter constraints, suggesting that single-line fits lack the depth needed to yield unique solutions. Although the LOC + MCMC approach allows for flexible modelling of molecular abundances and kinematics, it remains sensitive to prior assumptions and observational limitations. To overcome these limitations, future efforts should focus on multi-line fitting for each isotopologue.\

 A layered picture of the L1544 core was developed based on the morphology of the observed line profiles and modelling outcomes. Transitions showing deep self-absorption and/or extended profiles (e.g., HCO$^{+}$, H$_{2}$CO) trace the outer layers, while rarer isotopologues such as C$^{34}$S and HC$^{17}$O$^{+}$ reveal the kinematics of the inner regions through their double-peaked profiles. Among these, HC$^{17}$O$^{+}$ stands out as a probe of the innermost contracting layers, evidenced by its high velocity scaling and relatively constrained abundance drop-off radius.\
 
 A comparison of the fractional abundance profiles of the molecules obtained via LOC + MCMC models or by chemical modelling points to the problems of the chemical networks due to the uncertainties in elemental abundances input in chemical models, in particular sulphur. The poor reproduction of the sulphur chemistry hints at the need of the inclusion of a more consistent sulphur-depletion process, starting from the elemental cosmic abundance of S, to be able to model sulphur-bearing molecules towards pre-stellar cores. \ 

The results presented in this manuscript show that all of the molecular transitions trace different parts of the L1544 core with different physical, chemical and kinematic properties, revealing the sub-structure of the L1544 pre-stellar core. Future 3D models will be considered to assess whether the deviation from spherical symmetry is needed. Moreover, these results point to the need to use 3D radiative transfer modelling to be able to interpret increasingly sensitive and spectrally resolved spectra towards L1544.\

\textit{Acknowledgments.}
The authors gratefully acknowledge the support of the Max Planck Society. J. Ferrer Asensio thanks RIKEN Special Postdoctoral Researcher Program (Fellowships) for financial support. Moreover, the authors thank the anonymous referees for their constructive comments and helpful suggestions, which improved the clarity and quality of the manuscript. \

\newpage

\bibliography{AAS56751}{}
\bibliographystyle{aasjournal}

\newpage

\appendix
\section{Optical Depth Estimations}\label{optdepth}
\begin{figure}[H]
\centering
\includegraphics[width=9cm]{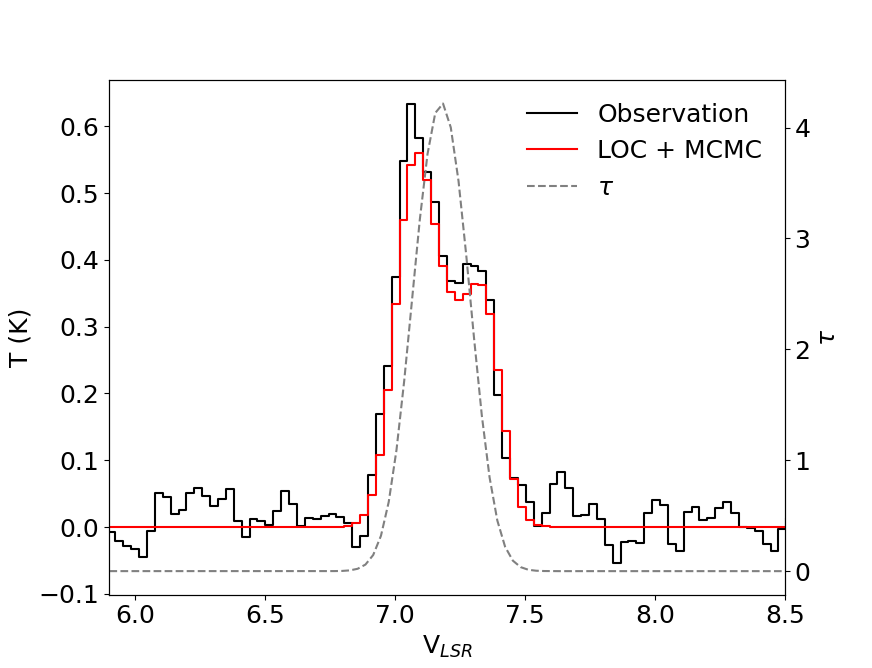}

\caption{The observed C$^{34}$S (2 - 1) transition is plotted in black, the LOC + MCMC spectra is plotted in red and the optical depth ($\tau$) is plotted with a grey dashed line. An additional y axis (right) indicating the optical depth is included. }
\label{tauc34sp}
\end{figure}

\begin{figure}[H]
\centering
\includegraphics[width=9cm]{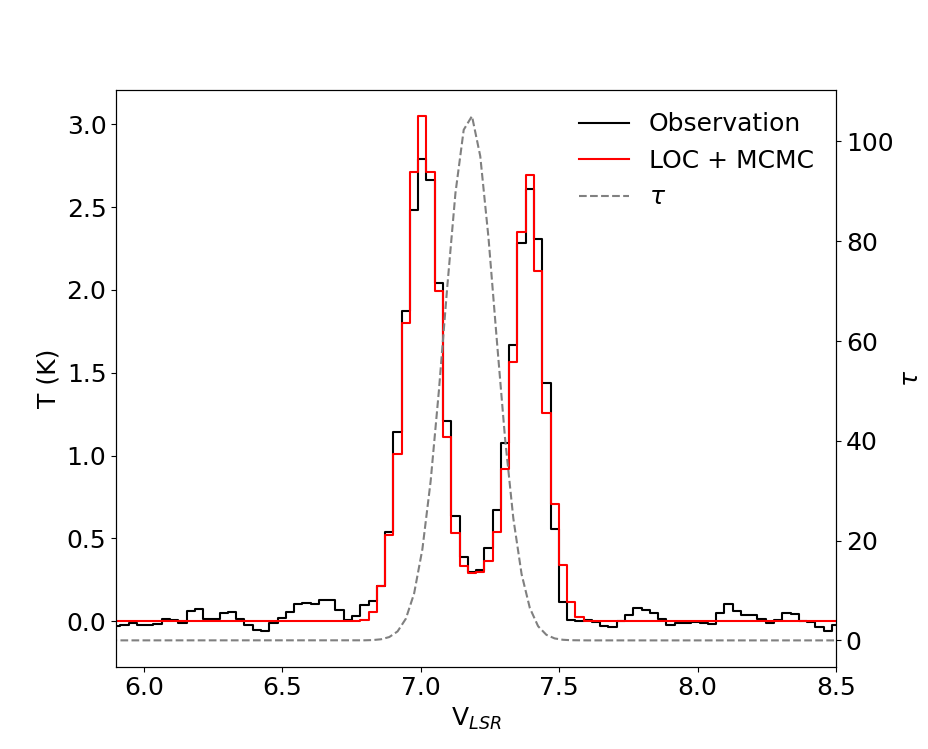}

\caption{Same as Figure \ref{tauc34sp}, but for the CS (2 - 1) transition.}
\label{taucsp}
\end{figure}

Estimations of the optical depth of selected transitions performed with the LOC software are presented. The optical depths are computed at different values of V$_{LSR}$ resulting into an optical depth profile. Figure \ref{tauc34sp} shows the C$^{34}$S (2 - 1) observed and LOC + MCMC modelled spectra (in black and red, respectively) alongside the optical depth profile (in dashed grey). The maximum optical depth of the line has a value of $\sim$  4.22. The optical depth profile appears double peaked.  Despite being from a rarer isotopologue, this line is sufficiently optically thick to present self absorption which, alongside the contraction of the core, results into the characteristic infall asymmetry profile. On the other hand, the optical depth profile of the CS (2 - 1) transition (dashed grey line in Figure \ref{taucsp}) appears single peaked, with a maximum optical depth value of $\sim$  105. \
 The maximum optical depth values computed from the LOC + MCMC for these two molecules are unexpectedly high. This is probably due to the high $a_{out}$ found by the LOC + MCMC approach. This result also stresses the fact that, with the lack of enough constraints, the result found by LOC + MCMC may not be the "correct" or a physically accurate result.

\newpage

\section{Corner Plots}

This Section introduces the corner plots for the LOC + MCMC modelling results presented in Section \ref{results3}. In all of the corner plots, the top left subplot corresponds to the inner abundance values explored in the modelling. The second column is the outer abundance. The third corresponds to the radius separating the abundance regions. The fourth shows the velocity profile scaling, and the fifth the turbulent velocity dispersion. The rest of the panels show the correlation between the different variables. On top of each column the resulting model parameter value and its errors are displayed. The model parameter values are additionally listed in the main text in Table \ref{modres}.\

\begin{figure}[H]
\centering
\includegraphics[width=8cm]{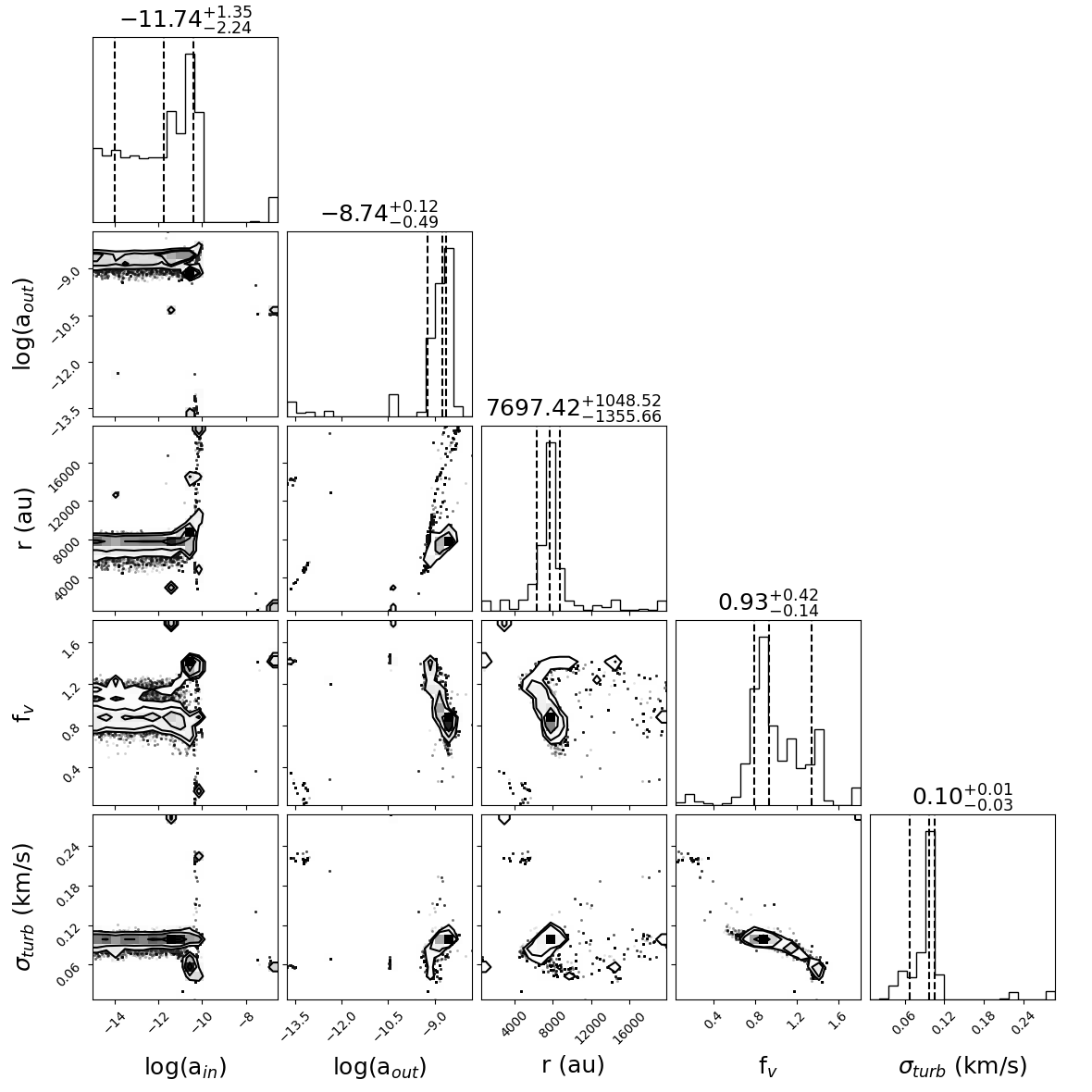}
\caption{Corner plot of the parameters used for the C$^{34}$S (2 - 1) LOC + MCMC in Figure \ref{c34sspec}. }
\label{c34scpp}
\end{figure}

\begin{figure}[H]
\centering
\includegraphics[width=8cm]{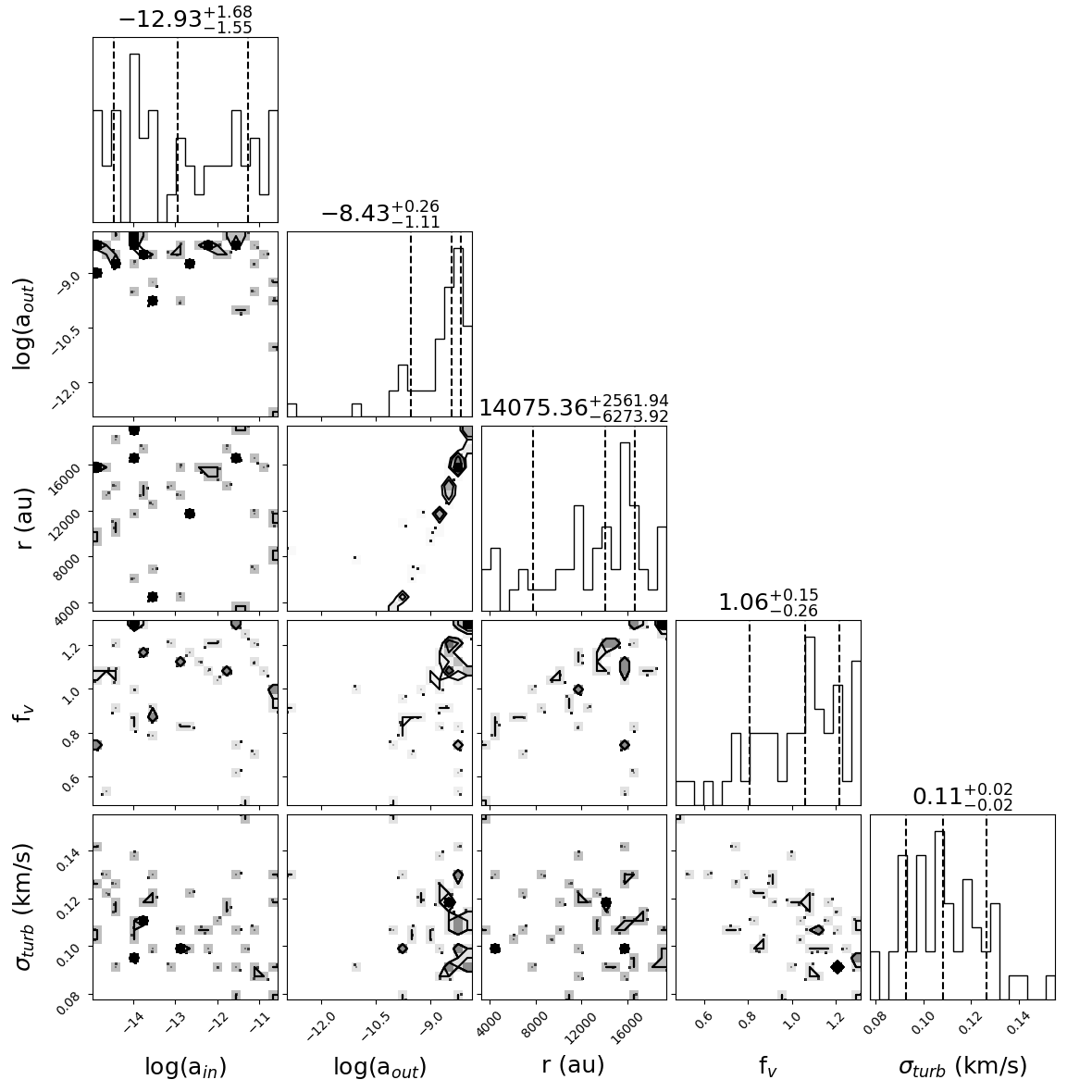}
\caption{Corner plot of the $^{34}$SO (2,3 - 1,2) parameters used for the LOC + MCMC in Figure \ref{34sospec}.}
\label{34socpp}
\end{figure}

\begin{figure}[H]
\centering
\includegraphics[width=8cm]{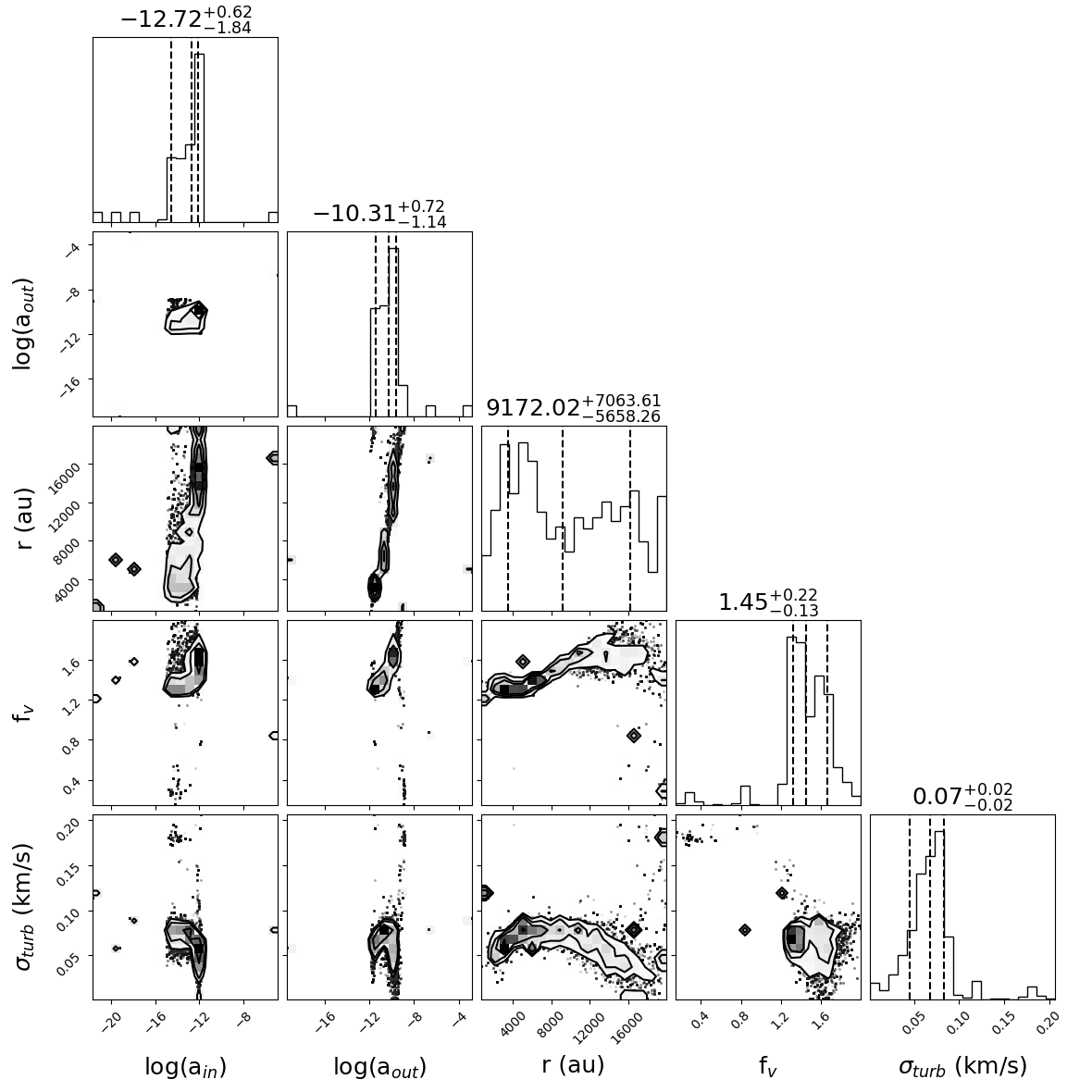}
\caption{Corner plot of the HC$^{17}$O$^{+}$ (1 - 0) parameters used for the LOC + MCMC in Figure \ref{hc17o+spec}.}
\label{hc17o+cpp}
\end{figure}

\begin{figure}[H]
\centering
\includegraphics[width=8cm]{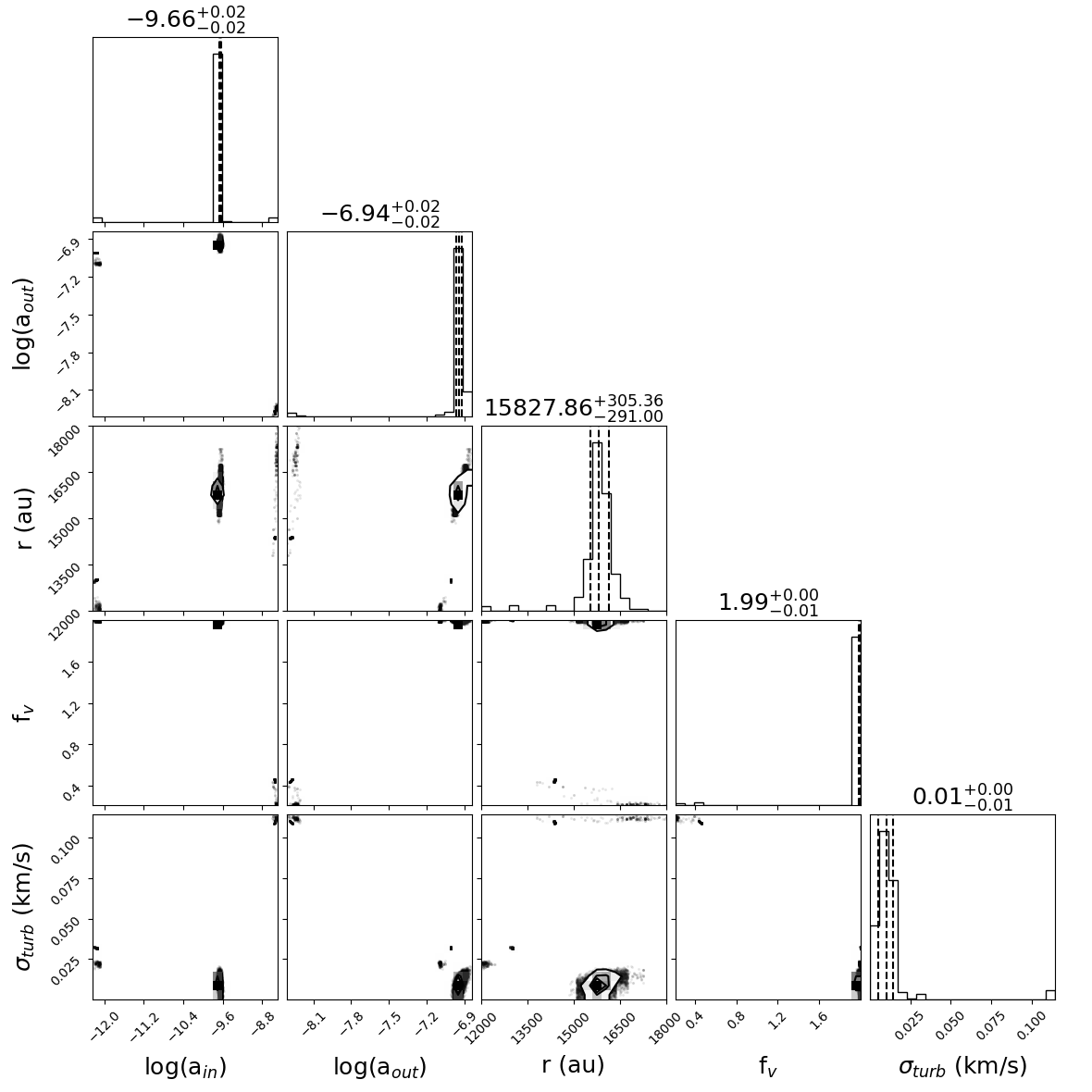}
\caption{Corner plot of the CS (2 - 1) parameters used for the LOC + MCMC in Figure \ref{csp}.}
\label{cscpp}
\end{figure}

\begin{figure}[H]
\centering
\includegraphics[width=8cm]{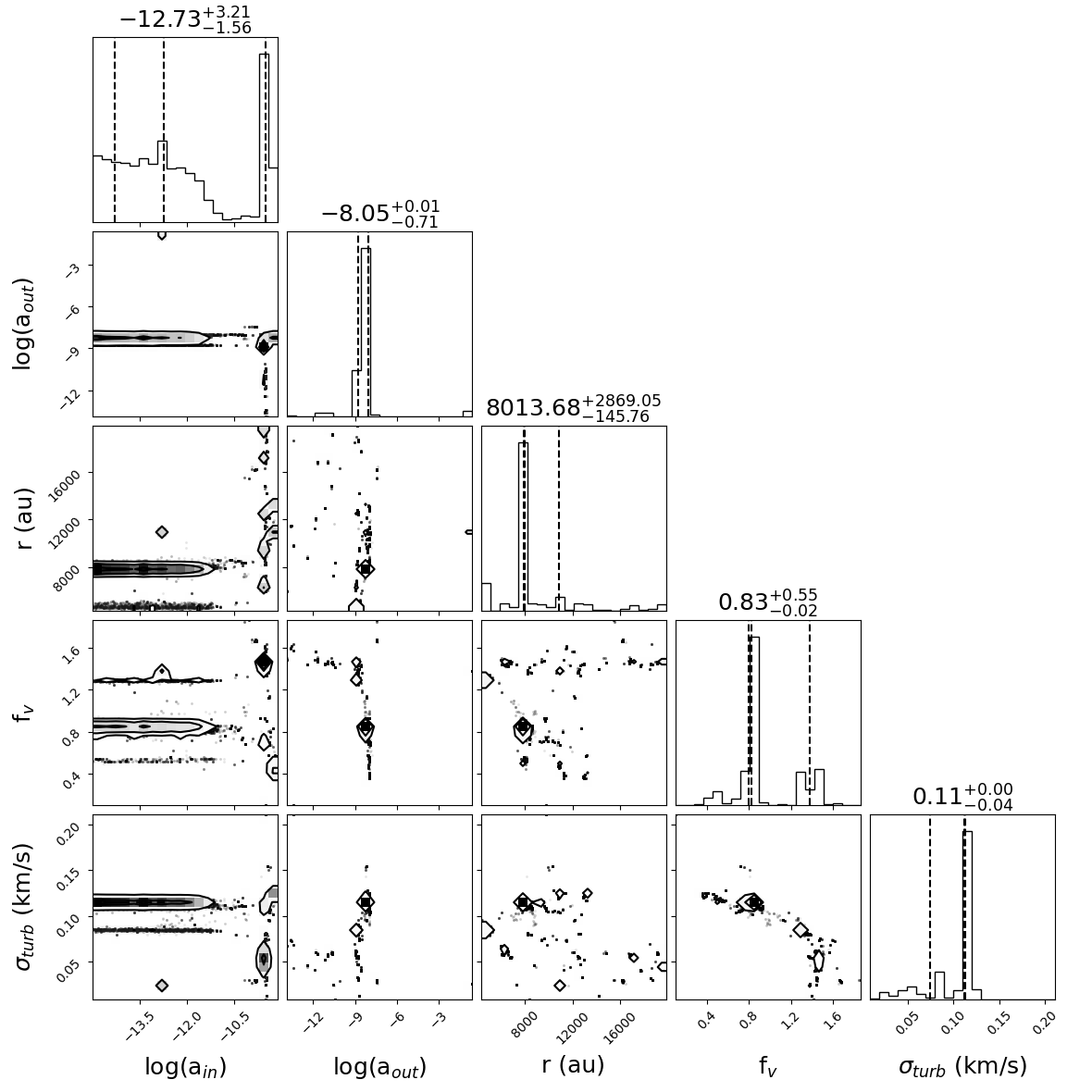}
\caption{Corner plot of the SO (2,3 - 1,2) parameters used for the LOC + MCMC in Figure \ref{sop}.}
\label{socpp}
\end{figure} 

\begin{figure}[H]
\centering
\includegraphics[width=8cm]{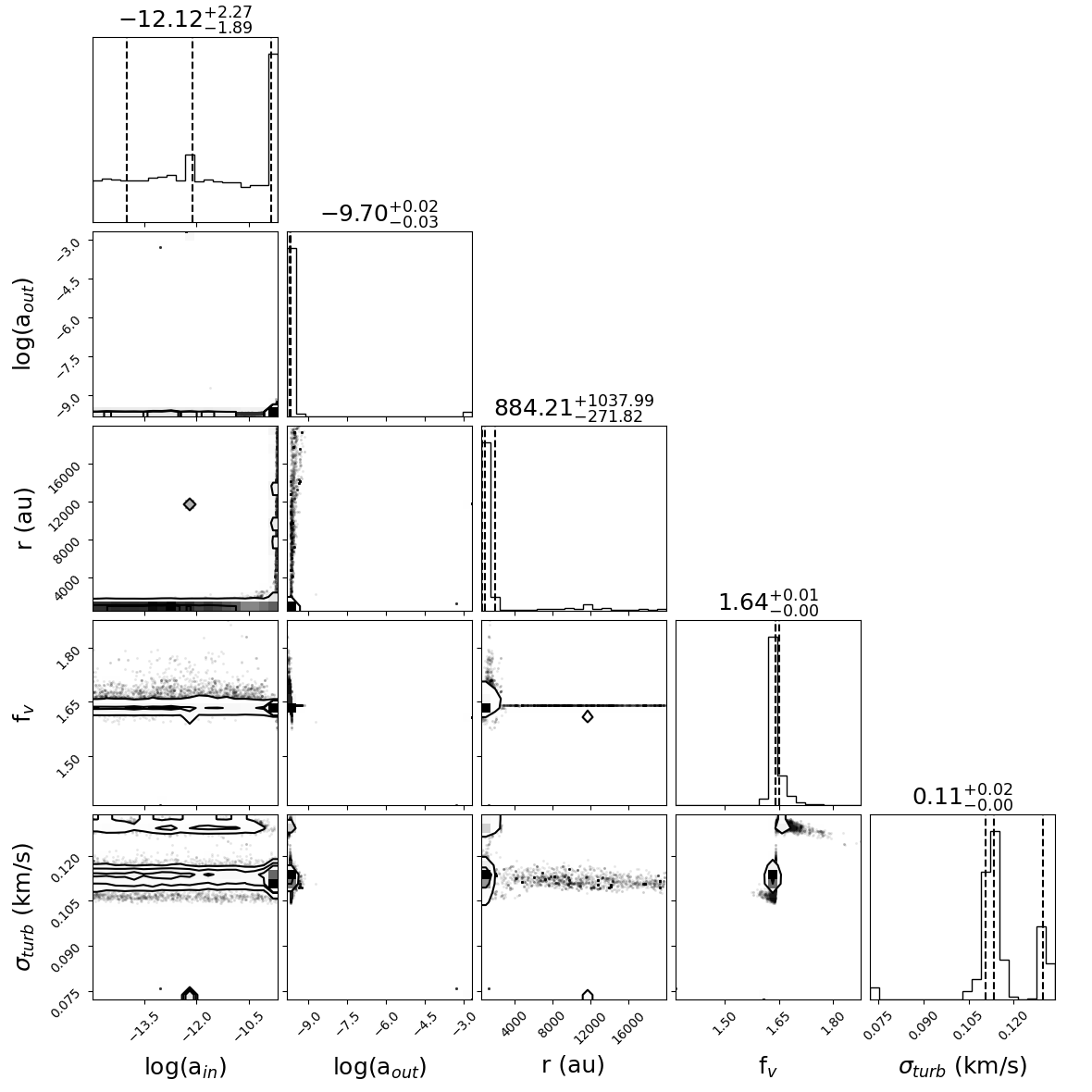}
\caption{Corner plot of the HCO$^{+}$ (1 - 0) parameters used for the LOC + MCMC in Figure \ref{hco+p}.}
\label{hco+cpp}
\end{figure} 

\begin{figure}[H]
\centering
\includegraphics[width=8cm]{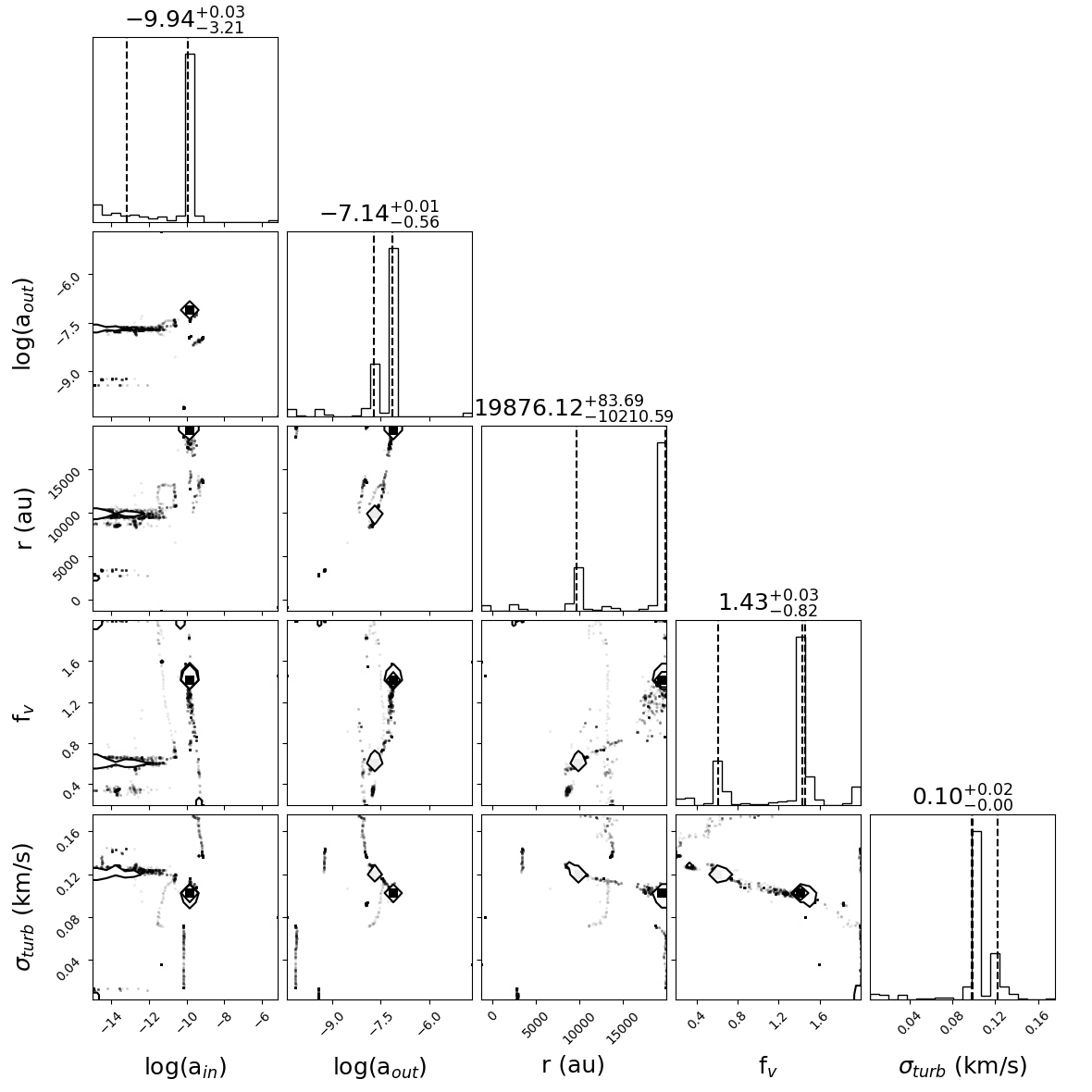}
\caption{Corner plot of the H$_{2}$CO (2$_{1,2}$ - 1$_{1,1}$) parameters used for the LOC + MCMC in Figure \ref{h2cop}.}
\label{h2cocpp}
\end{figure} 

\begin{figure}[H]
\centering
\includegraphics[width=8cm]{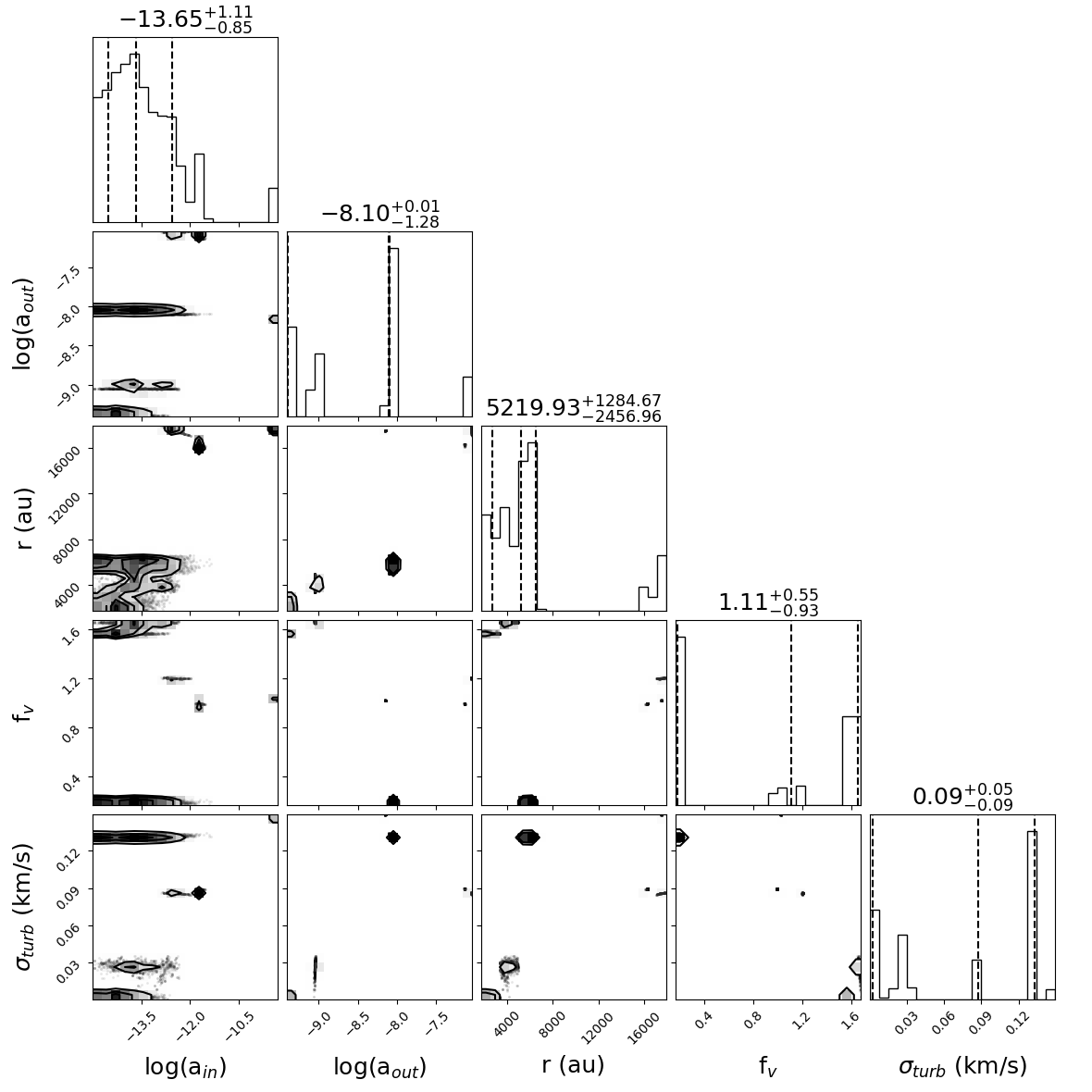}
\caption{Corner plot of the extended H$_{2}$CO (2$_{1,2}$ - 1$_{1,1}$) parameters used for the LOC + MCMC in Figure \ref{h2coep}.}
\label{h2coecpp}
\end{figure}

\begin{figure}[H]
\centering
\includegraphics[width=8cm]{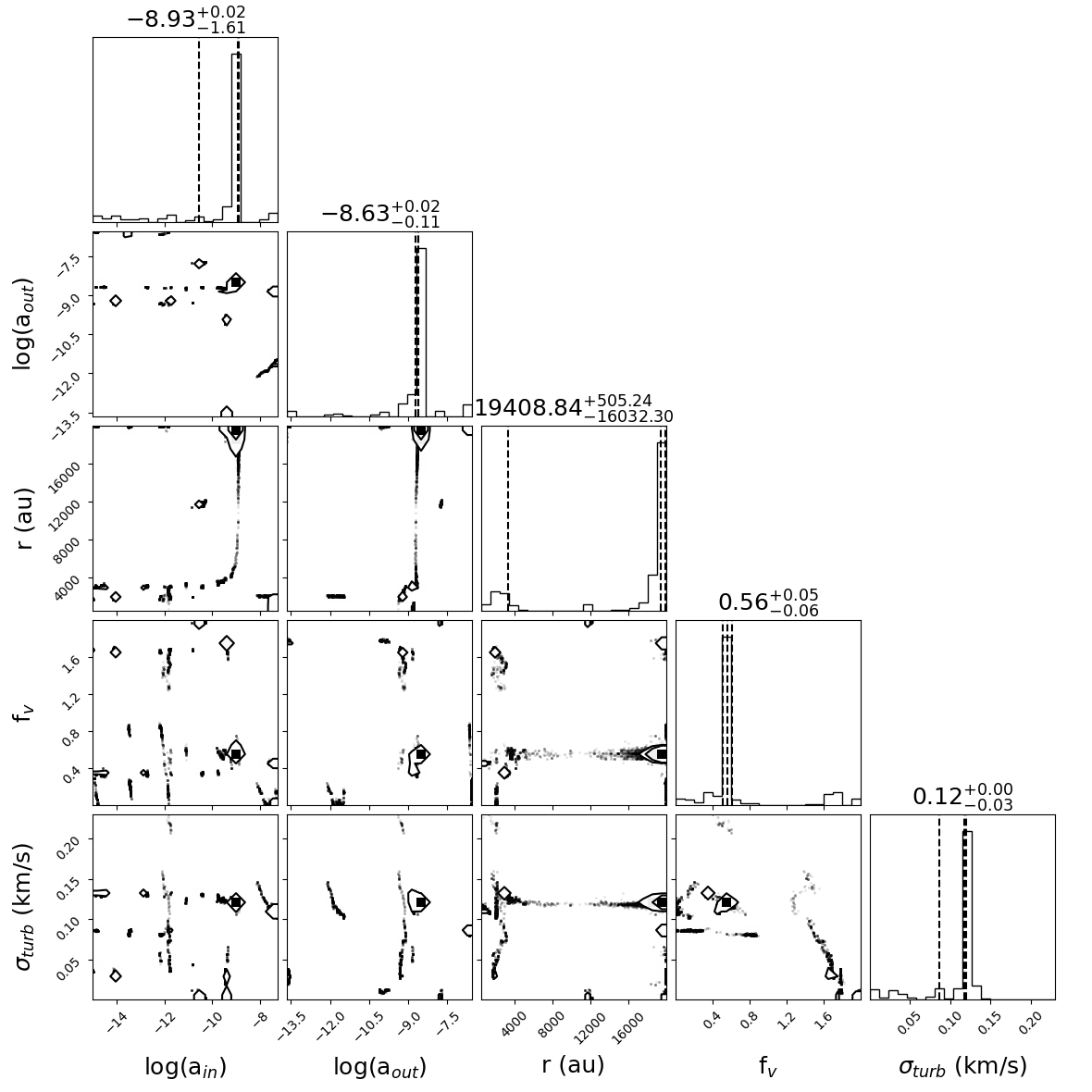}
\caption{Corner plot of the c-C$_{3}$H$_{2}$ (2$_{1,2}$ - 1$_{0,1}$) parameters used for the LOC + MCMC in Figure \ref{cc3h2p}.}
\label{cc3h2cpp}
\end{figure} 

\begin{figure}[H]
\centering
\includegraphics[width=8cm]{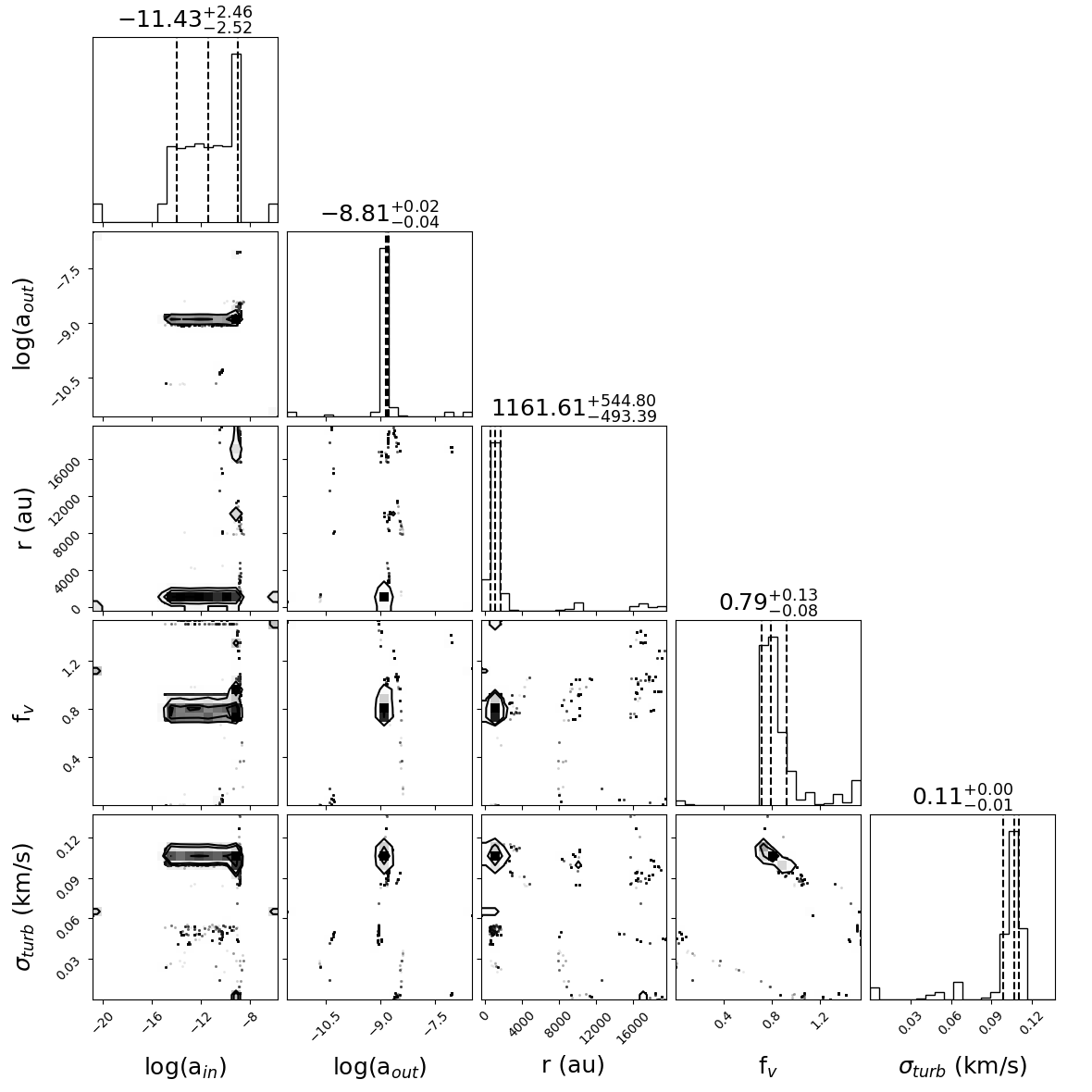}
\caption{Corner plot of the extended c-C$_{3}$H$_{2}$ (2$_{1,2}$ - 1$_{0,1}$) parameters used for the LOC + MCMC in Figure \ref{cc3h2ep}.}
\label{cc3h2ecpp}
\end{figure}

\newpage

\section{Additional LOC + MCMCs}\label{add}

Complementary models to those introduced in Section \ref{results3} are presented in the following Sections. The LOC + MCMC parameter values are summarised in Table \ref{modresa}.\

\begin{table*}[t]
\begin{center}
\caption{Additional LOC + MCMC Parameter Results}

\resizebox{\textwidth}{!}{\begin{tabular}{ lccccc } 
\hline\hline
Model & a$_{in}$ &  a$_{out}$ & $r$ & f$_{v}$ &  $\sigma_{turb}$\\
 & & & (au) &  & (km s$^{-1}$)\\
\hline
 CS + C$^{34}$S (2 - 1) &  6.92$^{+11.85}_{-0.48}$$\times$10$^{-9}$  &  1.44$^{+1.52}_{-1.34}$$\times$10$^{-7}$  &  9550.04$^{+1446.22}_{-1397.43}$  &  1.43$^{+0.01}_{-0.00}$  &  0.00$^{+0.00}_{-0.00}$   \\
SO + $^{34}$SO (2,3 - 1,2)  &  1.07$^{+138.90}_{-0.00}$$\times$10$^{-11}$  &  1.71$^{+1.79}_{-0.73}$$\times$10$^{-8}$  &  5567.16$^{+664.31}_{-2308.64}$  &  0.82$^{+1.17}_{-0.01}$  &  0.11$^{+0.00}_{-0.08}$ \\
HCO$^{+}$ + HC$^{17}$O$^{+}$ (1 - 0) extended  &  8.44$^{+107.90}_{-0.18}$$\times$10$^{-11}$  &  9.89$^{+19.29}_{-2.05}$$\times$10$^{-9}$  &  9428.14$^{+4493.90}_{-8456.55}$  &  0.87$^{+0.10}_{-0.19}$  &  0.11$^{+0.01}_{-0.01}$  \\

 \hline
\label{modresa}
\end{tabular}}

\tablecomments{The LOC + MCMC parameters and their errors are presented. For details on each of the models refer to the different subsections enclosed in the Appendix Section \ref{add}.}
\end{center}
\end{table*}

\subsection{CS + C$^{34}$S modelling}\label{csc34s}

In an attempt to improve the fit and parameter constraints presented in Section \ref{cs}, we construct a new model which fits the CS and C$^{34}$S (2 - 1) observed lines simultaneously. The idea behind this combined modelling is to have the same amount of variables explored with the MCMC with more observational constraints coming from two observed spectral lines instead of one. This model assumes that the main and rarer isotopologue share the same abundance profile, and the rarer isotopologue profile is scaled down by the $^{32}$S/$^{34}$S = 22 ratio \citep{wilson:94}. Thus, we have the same variables as for the CS and C$^{34}$S separate models: a$_{in}$, a$_{out}$, $r$, $f_{v}$ and  $\sigma_{turb}$. We test this combined modelling for both the non-extended and the extended models.\ 

\begin{figure}[H]
\centering
\includegraphics[width=7cm]{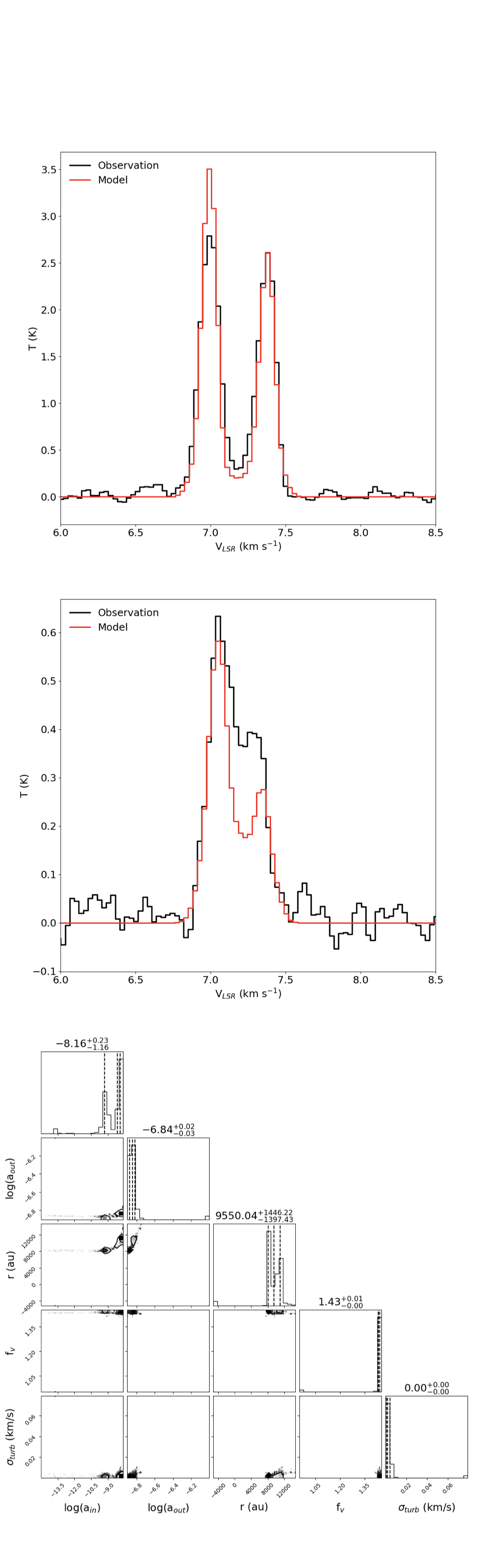}

\caption{\textit{Top panel:} CS (2 - 1) observations towards the L1544 dust peak in black and modelling results in red. \textit{Middle panel:} as the top panel but for C$^{34}$S (2 - 1). \textit{Bottom panel:} Corner plot for the CS + C$^{34}$S non-extended model.  }
\label{cspa}
\end{figure}

The combined non-extended model results for CS and C$^{34}$S can be found in Figure \ref{cspa}. The model fit parameters are presented in Table \ref{modresa} as well as in the corner plot in the bottom panel in Figure \ref{cspa}. The CS (2 - 1) modelled transition fits the observations worse than the one shown in Section \ref{cs}.  The model slightly over-predicts the intensity of the blue peak with respect to the observations (top panel, Figure \ref{cspa}). The C$^{34}$S (2 - 1) modelled transition fit is worse (Middle Panel, Figure \ref{cspa}) when compared to the individual C$^{34}$S model in Section \ref{c34s} {by underproducing the red part of the line and overproducing the self-absorption dip}.  The resulting model parameters, and specifically $a_{in}$, $a_{out}$ and  $\sigma_{turb}$ show values close to the ones found for the CS individual model. The combined non-extended model mainly fits CS, and thus, the modelled parameters are closer to the CS individual fit model ones.\ 

 Notice that the corner plot of this combined fit also resembles the corner plot for the CS individual fit with really narrow distributions. Moreover, the  $\sigma_{turb}$ value found in the combined fit of 0.00$^{+0.00}_{-0.00}$ also equals the prior distribution limit, being a sign that there may have been a problem with the exploration.

\subsection{SO + $^{34}$SO modelling}\label{so34so}

With the same aim as for CS + C$^{34}$S, we construct a new model to fit SO and $^{34}$SO (2,3 - 1,2) simultaneously. As for CS + C$^{34}$S (Section \ref{csc34s}) the model for SO and $^{34}$SO assumes that the main and rarer isotopologue share the same abundance profile but the rarer isotopologue profile is scaled down by the $^{32}$S/$^{34}$S = 22 ratio \citep{wilson:94}. As the SO line fits the observations with the non-extended model, we try the SO + $^{34}$SO combined fit with the non-extended model.\ 

\begin{figure}[h]
\centering
\includegraphics[width=7cm]{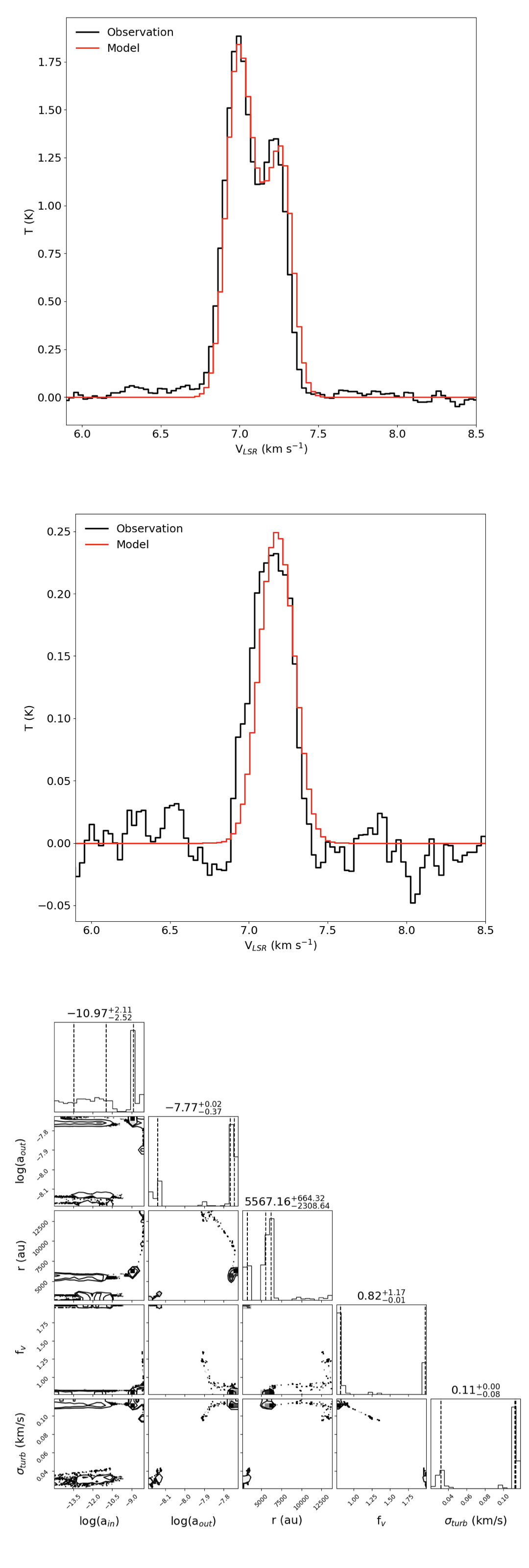}

\caption{\textit{Top panel:} SO (2,3 - 1,2) observations towards the L1544 dust peak in black and combined modelling results in red. \textit{Middle panel:} $^{34}$SO (2,3 - 1,2) observations towards the L1544 dust peak in black and combined modelling results in red. \textit{Bottom panel:} Corner plot of the parameters used for the SO + $^{34}$SO non-extended model. }
\label{sopp}
\end{figure}

The fit results can be found in Figure \ref{sopp}. The model parameters are presented in Table \ref{modresa} as well as in the corner plot in the bottom panel in Figure \ref{sopp}. The combined model fits SO and $^{34}$SO (Top and Middle Panels, respectively, Figure \ref{sopp}) similarly as the respective single models do (Figures \ref{sop} and \ref{34sospec}, respectively).  The $a_{in}$ and $a_{out}$ parameters are higher than for both the SO and $^{34}$SO individual fits. Both the $r$ and $f_{v}$ parameters are lower than the SO and $^{34}$SO individual models. Finally, the  $\sigma_{turb}$ value is the same than for the individual fits. \ 

\subsubsection{HCO$^{+}$ Redaelli et al. 2022 reproduction.}\label{hco+red}

\begin{figure}[h]
\centering
\includegraphics[width=7cm]{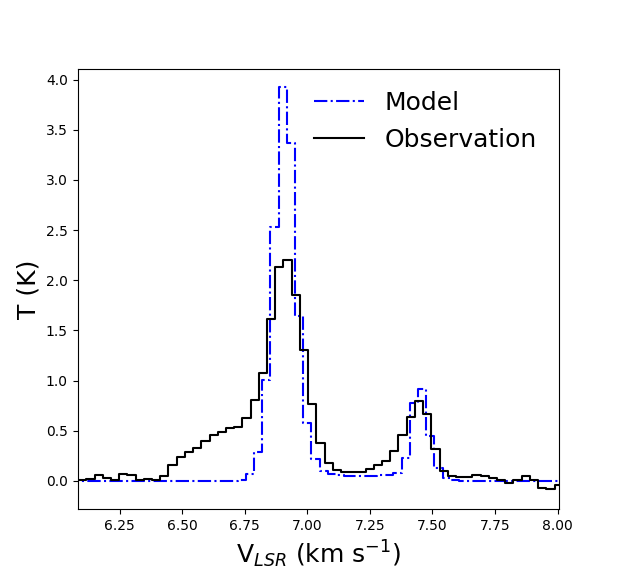}
\caption{The model using the approach of \cite{redaelli:22b} is shown in a dash-dotted blue line. Observations towards the L1544 dust peak are shown in black.}
\label{hco+e7}
\end{figure}

Using the same modelling approach as \cite{redaelli:22b}, which involves using the HCO$^{+}$ abundance profile from pyRate, we can reproduce their HCO$^{+}$ 1 - 0 transition fit (right panel in Figure 7, \citealp{redaelli:22b}), as shown in Figure \ref{hco+e7}. The model intensity increases by $\sim$14\% in our model.\ 

\subsubsection{HCO$^{+}$ + HC$^{17}$O$^{+}$ modelling}\label{hco+hc17o+}

The combined HCO$^{+}$ + HC$^{17}$O$^{+}$ (1 - 0) fit has been constructed similarly as the CS + C$^{34}$S and the SO + $^{34}$SO pair fits. The abundance profile used is the same as the one described in Section \ref{hco} for HCO$^{+}$ and it is scaled down by [$^{16}$O/$^{17}$O] = 2044 \citep{penzias:81, wilson:94} for HC$^{17}$O$^{+}$. As for the individual HCO$^{+}$ fit, we only test the extended fit.\ 

\begin{figure}[h]
\centering
\includegraphics[width=8cm]{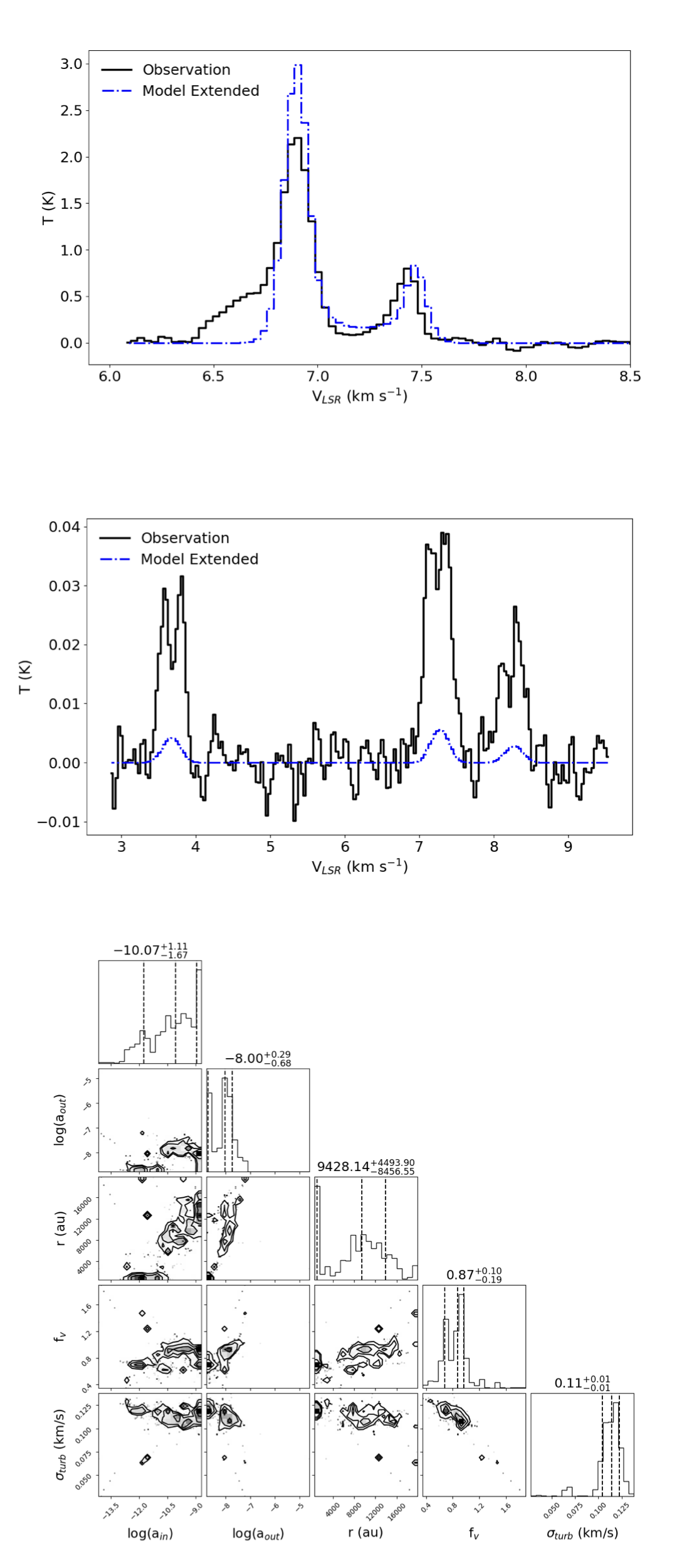}

\caption{\textit{Top panel:} HCO$^{+}$ (1 - 0) observations towards the L1544 dust peak in black and HCO$^{+}$ + HC$^{17}$O$^{+}$ extended modelling results in a dash-dotted blue line. \textit{Middle panel:} HC$^{17}$O$^{+}$ (1 - 0) observations towards the L1544 dust peak in black and combined modelling results in a dash-dotted blue line. \textit{Bottom panel:} Corner plot of the parameters used for the HCO$^{+}$ + HC$^{17}$O$^{+}$ extended model.  }
\label{hcopp}
\end{figure}

The HCO$^{+}$ + HC$^{17}$O$^{+}$ extended model fits are shown on the Top and Middle Panels, respectively, in Figure \ref{hcopp}. The model parameters are presented in Table \ref{modresa} as well as in the corner plot in the bottom panel in Figure \ref{hcopp}. The HCO$^{+}$ modelled spectrum does not reproduce the observed relative intensity of the blue and red components but reproduces the blue and red peak separation. The HC$^{17}$O$^{+}$ modelled spectrum does not fit the observed line intensity nor its profile. The HCO$^{+}$ + HC$^{17}$O$^{+}$ combined fit does not improve the constraints on the fractional abundance profile compared to the separate HCO$^{+}$ and HC$^{17}$O$^{+}$ fits.\

\end{document}